\documentclass[11pt]{amsart}

\usepackage{euler}
\usepackage{amscd} 
\usepackage{mathrsfs}
\usepackage{amsmath,amsbsy,amsthm,amsfonts,amssymb}
\usepackage{hyperref}
\theoremstyle{definition}

\numberwithin{equation}{section}

\def\x{\mathbf{x}}
\def\y{\mathbf{y}}

\def\R{\mathbb{R}}
\def\vac{|0\rangle}
\def\vacc{\langle 0|}
\def\aa{\alpha}
\def\bb{\beta}
\def\cc{\gamma}
\def\di{\varsigma}
\def\gg{b}
\def\Im{\mbox{Im}\,}
\def\Re{\mbox{Re}\,}

\title[Vacuum polarization with zero-range potentials on a hyperplane]{Vacuum polarization with zero-range potentials\\ on a hyperplane}

\author{Davide Fermi}
\address{Dipartimento di Matematica `Guido Castelnuovo', Universit\`a degli Studi di Roma `La Sapienza', Piazzale Aldo Moro 5, I-00185 Roma, Italy}
\email{fermidavide@gmail.com}
\urladdr{https://fermidavide.com}

\begin{document}

\begin{abstract} 
The quantum vacuum fluctuations of a neutral scalar field induced by background zero-range potentials concentrated on a flat hyperplane of co-dimension $1$ in $(d+1)$-dimensional Minkowski spacetime are investigated. Perfectly reflecting and semitransparent surfaces are both taken into account, making reference to the most general local, homogeneous and isotropic boundary conditions compatible with the unitarity of the quantum field theory. The renormalized vacuum polarization is computed for both zero and non-zero mass of the field, implementing a local version of the zeta regularization technique. The asymptotic behaviours of the vacuum polarization for small and large distances from the hyperplane are determined to leading order. It is shown that boundary divergences are soften in the specific case of a pure Dirac delta potential.
\end{abstract}

\maketitle

\begin{footnotesize}
\\
{\bf PACS 2010}: 03.70.+k; 11.10.Gh; 41.20.Cv; 02.30.Sa
\end{footnotesize}

\section{Introduction}

Casimir physics deals with the ubiquitous London-van der Waals dispersion forces, arising from the spontaneous polarization of neutral atoms and molecules, in a regime where retardation effects are not negligible. Accordingly, the resulting Casimir forces between macroscopic bodies are truly quantum and relativistic in nature.

In a pioneering work dating back to 1948 \cite{Cas48}, following a suggestion of Bohr, Hendrik Casimir made a groundbreaking theoretical prediction: two parallel, neutral conducting plates would experience a mutually attractive force $F = \hbar c\, {\pi^2 \over 240}\,{\Sigma/a^4}$ ($a$ and $\Sigma$ denoting, respectively, the distance between the plates and their surface area), due to a variation of the electromagnetic quantum vacuum energy induced by the presence of the plates themselves. This astonishing result indicates that a detailed microscopic description of the plates constituents is actually unnecessary for the computation of the previously mentioned dispersion forces, at least to leading order. Indeed, it is sufficient to consider effective models where relativistic quantum fields are influenced by classical boundaries, external potentials, or even curved or topologically non-trivial background geometries.
Building on this crucial feature, the study of the Casimir effect has nowadays become a well-established and extremely active line of research, both on the theoretical and on the experimental side. Here we content ourselves with mentioning the classical essays \cite{BKMM09,BMM01,DMRR11,KMM09,Milt01,MT97}, making also reference to the vast literature cited therein.

Assuming that quantum fields are confined by perfectly reflecting boundaries is a strong idealization: understandably, no real material is going to behave as a perfect conductor in any frequency range of the electromagnetic field. It comes as no surprise that a price must be paid for this simplification. As first pointed out by Deutsch and Candelas in 1979 \cite{DC79}, renormalized expectation values of local observables, such as the vacuum energy density, generically diverge in a non-integrable way as the boundary is approached. This leads inevitably to the emergence of anomalies in the computation of the associated global observables (see also \cite{BBLPRS15,FS98,KCD80}). Similar issues appear even if the confinement of the quantum field is produced by a smooth external potential diverging at infinity \cite{FP15}. On the contrary, no pathologies are expected to occur when the external potential is regular and vanishes rapidly enough at large distances.

An intermediate regime between smooth confining potentials and hard boundaries can be realized through singular zero-range potentials. Their mathematical description ultimately amounts to prescribing suitable boundary conditions for the quantum field on sets of small co-dimension ($1,2$ or $3$), where the distributional potentials are supposed to be concentrated. At the same time, such singular potentials can often be interpreted as limits (in resolvent sense) of sharply peaked, regular potentials. More technical details on these subjects can be found, e.g., in \cite{AGHH05,AK99,NGMMR20,Po01}. Nowadays, a quite rich literature is available regarding the analysis of Casimir-type settings with external zero-range potentials.
The Casimir effect in presence of surface Dirac delta potentials, interpreted as semi-transparent walls responsible for a partial confinement of the quantum field, was first addressed by Mamaev and Trunov \cite{MT81} and later examined in various configurations by several authors \cite{BBK,BHR92,CMK07,FL08,GJK02,K06,M04,MMM13}. More recently, considerable attention was devoted to the study of renormalized vacuum expectations of global observables (such as the total energy) in presence of generalized zero-range interactions concentrated on sets of co-dimension 1, corresponding to mixtures of $\delta-\delta'$ potentials \cite{AM13,BMS19,BSA16,CFP17,MM15,MSDT20}. Before proceeding, let us also mention that various models with point impurities, modelled via distributional potentials concentrated on sets of co-dimension 3, were analysed in \cite{ACSZ10,ACS16,BM15,BP17,FPSym,Grats18,Grats19,Sca05,SZ09}.

The present work studies the vacuum fluctuations of a canonically quantized, neutral scalar field in $(d+1)$-dimensional Minkowski spacetime (with $d \geqslant 1$) in presence of a flat hyperplane of co-dimension $1$. Both the massive and massless theories are considered. The presence of the hyperplane is described in terms of boundary conditions for the field and its normal derivative. It is worth remarking that all local, homogeneous and isotropic boundary conditions compatible with the unitarity of the quantum field theory are taken into account. Of course, two qualitatively different scenarios are allowed. The first one corresponds to a perfectly reflecting plane, yielding a total confinement of the field on either of the half-spaces that it separates; this setting is naturally portrayed in terms of classical boundary conditions of Dirichlet, Neumann or Robin type. The second one refers to a semitransparent plane, which can be tunnelled through by the quantum field; this situation is described making reference to generalized $\delta$-$\delta'$ potentials concentrated on the plane.

The main object of investigation is the vacuum polarization, namely the renormalized expectation value of the field squared at any spacetime point. This is computed implementing the $\zeta$-regularization technique in the formulation outlined in \cite{WSB} (see also \cite{FP11,FP16,DF20}), which allows to derive explicit integral representations in all cases of interest. These representations are then employed to determine the asymptotic behaviour of the vacuum polarization close to the hyperplane and far away from it. In this connection, the primary purpose is to inspect the presence of boundary divergences. For a perfectly reflecting hyperplane, it is found that the vacuum polarization always diverges near the plane (logarithmically for $d = 1$ and with a power law for $d \geqslant 2$, with respect to the distance from the plane); notably, the leading order term in the asymptotic expansion is always independent of the parameters describing specific boundary conditions. Similar divergences also occur for a semitransparent plane; however in this case the leading order asymptotics depend explicitly on the parameters appearing in the characterization of the boundary conditions. To say more, the leading order divergent contribution is absent for a specific choice of the parameters, corresponding to a pure Dirac delta potential. Some motivations explaining why this very model plays a somehow distinguished role are presented.

The paper is organized as follows. Section \ref{sec:gen} provides an overview of the local zeta regularization framework described in \cite{WSB}. In Section \ref{sec:perfect} the renormalized vacuum polarization for a scalar field in presence of a perfectly reflecting plane is analysed. The analogous observable in the case of a semitransparent plane is examined in Section \ref{sec:semitr}. In both Sections \ref{sec:perfect} and \ref{sec:semitr} the case of a massive field is first considered, and the corresponding massless theory is subsequently addressed by a limiting procedure. Finally, Appendix \ref{app:heatperf} presents a self-contained derivation of the heat kernel on the half-line for generic Robin boundary conditions at the origin, a tool used in the computations of Section \ref{sec:perfect}.

\section{General theory}\label{sec:gen}

The purpose of this section is to present a brief and self-contained summary of some general techniques extracted from \cite{WSB}, to be systematically employed in the sequel.

\subsection{Quantum field theory and the fundamental operator.}\label{subsec:QFTA}

We work in natural units of measure ($c = 1$, $\hbar = 1$) and identify $(d + 1)$-dimensional Minkowski spacetime with $\R^{d+1}$ using a set of inertial coordinates $(x^{\mu})_{\mu \,=\, 0,1,\,...\,,\,d} \equiv (t,\x)$, such that the Minkowski metric has components $(\eta_{\mu\nu}) = \mbox{diag}\{-1,1,...,1\}$.

Making reference to the standard formalism of canonical quantization, we describe a neutral scalar field living on a spatial domain $\Omega \subset \R^{d}$ as an operator-valued distribution $\hat{\phi} : (t,\x) \in \R \times \Omega \mapsto \hat{\phi}(t,\x) \in \mathcal{L}_{sa}(\mathfrak{F})$. Here $\mathfrak{F}$ is the bosonic Fock space constructed on the single-particle Hilbert space $L^2(\Omega)$ of square-integrable functions, and $\mathcal{L}_{sa}(\mathfrak{F})$ is the set of unbounded self-adjoint operators on it. We denote with $\vac \in \mathfrak{F}$ the corresponding vacuum state (not to be confused with the true Minkowskian vacuum) and assume that the dynamics is determined by a generalized Klein-Gordon equation of the form
\begin{equation*}
(\partial_{tt} + \mathcal{A}) \hat{\phi} = 0\,,
\end{equation*}
where $\mathcal{A} : \mbox{dom}(\mathcal{A}) \subset L^2(\Omega) \to L^2(\Omega)$ is a non-negative and self-adjoint operator on the single-particle Hilbert space. The non-negativity of $\mathcal{A}$ is in fact an indispensable requirement for a well-behaved quantum field theory, free of pernicious instabilities.
In typical applications, $\mathcal{A}$ is a Schr\"odinger-type differential operator on the spatial domain $\Omega$, possibly including a static external potential $V: \Omega \to \R$, \textit{i.e.},
\begin{equation*}
\mathcal{A} = - \Delta + V\,.
\end{equation*}

Correspondingly, whenever the spatial domain has a boundary $\partial \Omega$ it is essential to specify suitable conditions on it. We understand these boundary conditions to be encoded in the definition of the operator self-adjointness domain $\mbox{dom}(\mathcal{A})$. It goes without saying that the class of admissible potentials and boundary conditions is restricted by the fundamental hypotheses of self-adjointness and non-negativity for $\mathcal{A}$.

The configurations analysed in the present work regard a scalar field influenced solely by the presence of an either perfectly reflecting or semitransparent hyperplane $\pi$ which, without loss of generality, can be parametrized as
\begin{equation}\label{eq:pi}
\pi = \{\x \equiv (x_1,\dots,x_d) \in \R^{d}\,|\,x_1 = 0\}\,.
\end{equation}

As already mentioned in the Introduction, the coupling between the field and this hyperplane can always be described in terms of suitable boundary conditions for the field and its normal derivative on $\pi$. Accordingly, in all cases we shall characterize the fundamental operator $\mathcal{A}$ as a self-adjoint extension of the closable symmetric operator $(- \Delta + m^2) \!\upharpoonright\! C^{\infty}_c(\R^d \setminus \pi)$ on $L^2(\R^d \setminus \pi) \equiv L^2(\R^d)$ (here $m = const. \geqslant 0$ indicates the mass of the field). We refer to subsection \ref{subsec:fact} for more details.

\subsection{$\zeta$-regularization and renormalization}
As well known, a quantum field theory of the type outlined in the preceding subsection is typically plagued by ultraviolet divergences. A viable way to cure these divergences is the $\zeta$-regularization approach described in \cite{WSB}. Following this approach and assuming for now that $\mathcal{A}$ is strictly positive,\footnote{With this we mean that the spectrum $\sigma(\mathcal{A})$ is contained in $[\varepsilon,+\infty)$ for some $\varepsilon > 0$. In the sequel we will indicate how to relax this condition.} we firstly introduce the \textit{$\zeta$-smeared field operator}
\begin{equation*}
\hat{\phi}^{u} := (\mathcal{A}/\kappa^2)^{-u/4}\, \hat{\phi}\,;
\end{equation*}
here $u \in \mathbb{C}$ is the regularizing parameter and $\kappa > 0$ is a mass scale factor, included for dimensional reasons. Notice that the initial non-regularized theory is formally recovered setting $u = 0$.

Next, we consider the \textit{$\zeta$-regularized vacuum polarization} at any spacetime point $(t,\x) \in \R \times \Omega$, that is the regularized 2-point function at equal times evaluated along the space diagonal $\{\x,\y \in \Omega\,|\,\y = \x\}$:
\begin{equation*}
\vacc \big(\hat{\phi}^{u}(t,\x)\big)^2 \vac \equiv \vacc \hat{\phi}^{u}(t,\x) \hat{\phi}^{u}(t,\y) \vac \Big|_{\y = \x}\,.
\end{equation*}
This quantity can be expressed in terms of the integral kernel associated to a suitable complex power of the fundamental operator $\mathcal{A}$; more precisely, we have \cite[Eq. (2.26)]{WSB}
\begin{equation}\label{eq:fiuA}
\vacc \big(\hat{\phi}^{u}(t,\x)\big)^2 \vac = {\kappa^u \over 2}\,\mathcal{A}^{-{u+1 \over 2}}(\x,\y)\Big|_{\y = \x}\,.
\end{equation} 
Notice that the expression on the right-hand side of \eqref{eq:fiuA} does not depend on the time coordinate $t \in \mathbb{R}$, as expected for static configurations like the ones we are considering.

On very general grounds it can be shown that the function $(\x,\y) \mapsto \mathcal{A}^{-{u+1 \over 2}}(\x,\y)$ belongs to $C^{j}(\Omega \times \Omega)$ for any $u \in \mathbb{C}$ and $j \in \{1,2,3,\dots\}$ such that $\Re u > d - 1 + j$. Especially, let us remark that the said function is regular along the diagonal $\{\y = \x\}$ for $\Re u$ large enough. Furthermore, for any fixed $\x,\y \in \Omega$ (even for $\y = \x$), the map $u \mapsto \mathcal{A}^{-{u+1 \over 2}}(\x,\y)$ is analytic in the complex strip $\{u \in \mathbb{C}\,|\,\Re u > d-1\}$ and possesses a meromorphic extension to the whole complex plane $\mathbb{C}$ with at most simple pole singularities \cite{DFT,WSB,Minak,Seeley}.
In light of these results, we proceed to define the \textit{renormalized vacuum polarization} at $(t,\x) \in \R \times \Omega$ as
\begin{equation}\label{eq:fi2ren}
\vacc \hat{\phi}^{2}(t,\x) \vac_{ren} := RP\Big|_{u = 0} \vacc \big(\hat{\phi}^{u}(t,\x)\big)^2 \vac\,.
\end{equation}
Here and in the following we denote with $RP\big|_{u = 0}$ the regular part of the Laurent expansion near $u = 0$.\footnote{For any complex-valued meromorphic function $f$ defined in a complex neighbourhood of $u = 0$, making reference to its Laurent expansion $f(u) = \sum_{\ell = -\infty}^{+\infty} f_\ell\,u^\ell$ we define the regular part as $(RP\, f)(u) = \sum_{\ell = 0}^{+\infty} f_\ell\,u^\ell$, which yields in particular $RP\big|_{u = 0}\, f = f_0$.} Notably, if no pole arises at $u = 0$, Eq. \eqref{eq:fi2ren} simply amounts to evaluating the analytic continuation at this point and ultraviolet renormalization is attained with no need to subtract divergent quantities; on the contrary, when the meromorphic extension has a pole at $u = 0$, Eq. \eqref{eq:fi2ren} matches a minimal subtraction prescription \cite{BVW,WSB,WaldZ}.

Before we proceed, let us point out that a modification of the above construction is required whenever the fundamental operator $\mathcal{A}$ is non-negative but not strictly positive, namely, when its spectrum contains a right neighbourhood of $0$. In this case an infrared cut-off must be added in advance, and ultimately removed after renormalization of ultraviolet divergences. For example, one can replace $\mathcal{A}$ with $\mathcal{A} + m^2$ ($m > 0$) and compute the limit $m \to 0^{+}$ at last, after analytic continuation at $u = 0$. Concerning the present work, this modification plays a key role when the field is massless ($m = 0$); in this case the renormalized vacuum polarization is determined as the zero mass limit ($m \to 0^{+}$) of the analogous quantity in the massive theory, namely,
\begin{equation}\label{eq:fi2renm0}
\vacc \hat{\phi}^{2}(t,\x) \vac_{ren}^{(massless)} := \lim_{m \to 0^{+}} \vacc \hat{\phi}^{2}(t,\x) \vac_{ren}^{(massive)}\,.
\end{equation}

\subsection{Factorized configurations}\label{subsec:fact}

Let us now restrict the attention to product configurations with
$$ \Omega = \Omega_1 \times \mathbb{R}^{d-1} \,, \qquad \mathcal{A} = \mathcal{A}_1 \otimes \mathbf{1}_{d-1} + \mathbf{1}_{1} \otimes (-\Delta_{d-1})\,, $$ 
where $\Omega_1 \subset \mathbb{R}$ is any open interval, $\mathcal{A}_1$ is a positive self-adjoint operator on $L^2(\Omega_1)$, and $-\Delta_{d-1}$ indicates the free Laplacian on $\mathbb{R}^{d-1}$ with $\mbox{dom}(-\Delta_{d-1}) = H^2(\mathbb{R}^{d-1}) \subset L^2(\mathbb{R}^{d-1})$. It is implied that $\mbox{dom}(\mathcal{A}) = \overline{\mbox{dom}(\mathcal{A}_1) \!\otimes\! H^2(\mathbb{R}^{d-1})} \subset L^2(\Omega) \equiv L^2(\Omega_1) \!\otimes\! L^2(\mathbb{R}^{d-1})$.

Under these circumstances, everything is determined upon factorization by the reduced operator $\mathcal{A}_1$ acting on the 1-dimensional spatial domain $\Omega_1$. In particular, let us highlight that for any $\tau> 0$ the heat kernel $e^{-\tau \mathcal{A}}(\x,\y)$ and the reduced analogue $e^{-\tau \mathcal{A}_1}(x_1,y_1)$ fulfil
\begin{equation*}
e^{-\tau \mathcal{A}}(\x,\y) = e^{-\tau \mathcal{A}_1}(x_1,y_1)\; e^{\tau \Delta_{d-1}}(\x_{d-1},\y_{d-1})\,,
\end{equation*}
where we put $\x_{d-1} \equiv (x_2,\dots,x_d) \in \mathbb{R}^{d-1}$ and denoted with $e^{\tau \Delta_{d-1}}(\x_{d-1},\y_{d-1})$ the free heat kernel in $\mathbb{R}^{d-1}$, that is
\begin{equation*}
e^{\tau \Delta_{d-1}}(\x_{d-1},\y_{d-1}) = {1 \over (4\pi \tau)^{d-1 \over 2}}\;e^{-{|\x_{d-1}-\,\y_{d-1}|^2 \over 4\tau}}\,.
\end{equation*}

Taking into account the above considerations, from the basic Mellin-type identity
\begin{equation*}
\mathcal{A}^{-s}(\x,\y) = {1 \over \Gamma(s)} \int_{0}^{\infty} d\tau\;\tau^{s-1}\,e^{-\tau \mathcal{A}}(\x,\y)\,,
\end{equation*}
we infer by direct evaluation
\begin{equation*}
\mathcal{A}^{-s}(\x,\y)\Big|_{\y = \x} = {1 \over (4\pi)^{d-1 \over 2}\, \Gamma(s)} \int_{0}^{\infty} d\tau\;\tau^{s-{d+1 \over 2}}\,e^{-\tau \mathcal{A}_1}(x_1,y_1)\Big|_{y_1 = x_1}\,.
\end{equation*}
Together with Eq.\,\eqref{eq:fiuA}, the latter relation allows us to derive the following representation formula for the $\zeta$-regularized vacuum polarization, valid for $u \in \mathbb{C}$ with $\Re u > d - 1$:
\begin{equation}\label{eq:fiuheat}
\vacc \big(\hat{\phi}^{u}(t,\x)\big)^2 \vac 
= {\kappa^u \over 2\,(4\pi)^{d-1 \over 2}\, \Gamma({u+1 \over 2})} \int_{0}^{\infty} d\tau\;\tau^{{u - d \over 2}}\,e^{-\tau \mathcal{A}_1}(x_1,y_1)\Big|_{y_1 = x_1}\,.
\end{equation}

Let us now return to the configurations portrayed at the end of subsection \ref{subsec:QFTA}, involving a scalar field restrained by the presence of a hyperplane $\pi$. Whenever the effective interaction between the field and $\pi$ is isotropic and homogeneous along the hyperplane, these configurations exhibit the factorization property discussed above. More precisely, referring to the parametrization \eqref{eq:pi} of $\pi$, under the hypotheses just mentioned it is natural to consider the reduced domain $\Omega_1 = \mathbb{R} \setminus \{0\}$ and to characterize the associated operator $\mathcal{A}_1$ as a self-adjoint extension of the symmetric operator $(-\,\partial_{x_1 x_1}\!+m^2) \upharpoonright C^{\infty}_c(\mathbb{R} \setminus \{0\})$.\linebreak
We recall that the domains of these self-adjoint extensions are indeed restrictions of the maximal domain $H^2(\mathbb{R} \setminus \{0\}) \equiv H^2(\mathbb{R}_{-}) \oplus H^2(\mathbb{R}_{+})$ (where $\mathbb{R}_{+} \equiv (0,+\infty)$ and $\mathbb{R}_{-} \equiv (-\infty,0)$), determined by suitable boundary conditions at the gap point $x_1 = 0$.
As a matter of fact, the models discussed in the upcoming Sections \ref{sec:perfect} and \ref{sec:semitr} encompass all admissible self-adjoint realizations of the reduced operator $-\,\partial_{x_1 x_1}\!+m^2$ on $\mathbb{R}\setminus \{0\}$, respecting the basic requirement of positivity. 
Notice however that this scheme does not reproduce the entire class of (positive) self-ajoint realizations of the full operator $-\Delta + m^2$ on $\mathbb{R}^{d} \setminus \pi$, since non-homogeneous and non-local self-adjoint realizations are being omitted.\footnote{Let us mention a pair of examples which are not covered by our analysis. On one hand, local but non-homogeneous boundary conditions appear in the description of $\delta$-potentials supported on $\pi$ with non-constant coupling coefficients, formally corresponding to operators like  $-\Delta + m^2 + \alpha(\x_{d-1})\, \delta_{\pi}$ (with $\x_{d-1} \equiv (x_2,\dots,x_d) \in \mathbb{R}^{d-1} \simeq \pi$). On the other hand, homogeneous but non-local boundary conditions are used to characterize operators like $-\Delta + m^2 + \alpha(-\Delta_{d-1})\, \delta_{\pi}$, with $\alpha(-\Delta_{d-1})$ a suitable self-adjoint operator on $L^2(\pi)$ defined by functional calculus \cite{CFP17}.}

In the following sections, after providing a precise definition of the reduced operator $\mathcal{A}_1$ under analysis and an explicit expression for the associated heat kernel $e^{-\tau \mathcal{A}_1}(x_1,y_1)$, we proceed to construct the analytic continuation of the map $u \mapsto \vacc \big(\hat{\phi}^{u}(t,\x)\big)^2 \vac$ to the whole complex plane starting from the representation formula \eqref{eq:fiuheat}. The renormalized vacuum fluctuation is ultimately computed following the prescriptions \eqref{eq:fi2ren} and \eqref{eq:fi2renm0}.

\section{Perfectly reflecting plane}\label{sec:perfect}

In this section we analyse the admissible scenarios where the hyperplane $\pi$ behaves as a perfectly reflecting surface, providing a total decoupling of the two half-spaces which it separates. To this purpose, taking into account the general arguments presented in the preceding Section \ref{sec:gen} and making reference to \cite[Thm. 3.2.3]{AK99}, we consider the family of reduced operators labelled as follows by the elements $\mathbf{h}^{\pm} \!= (h_0^{\pm}, h_1^{\pm})$ of the real projective space $\mathbf{P}_1$:
\begin{eqnarray}
& \mbox{dom}(\mathcal{A}_1) := \big\{\psi \in H^2(\mathbb{R} \!\setminus\! \{0\})\,\big|\, h_0^{+} \psi'(0^{+}) = h_1^{+} \psi(0^{+}),\, h_0^{-} \psi'(0^{-}) = h_1^{-} \psi(0^{-})\big\}\,, \nonumber\\
& \mathcal{A}_1\, \psi = (-\,\partial_{x_1 x_1}\! + m^2)\, \psi \quad \mbox{in\; $\mathbb{R} \!\setminus\! \{0\}$}\,.\label{eq:Aperf}
\end{eqnarray}
Let us remark that the above definition of $\mbox{dom}(\mathcal{A}_1)$ entails classical boundary conditions of Neumann, Dirichlet or Robin type, chosen independently on the two sides of the gap point $x_1 = 0$ (\textit{viz.}, on the two sides of the hyperplane $\pi$). Especially, Neumann conditions correspond to $h_1^{\pm} = 0$ ($h_0^{\pm} \neq 0$) and Dirichlet ones to $h_0^{\pm} = 0$ ($h_1^{\pm} \neq 0$). In passing, we also mention that the Casimir effect for Robin boundary conditions was previously analysed in \cite{EOS09,RS02,SAD06}.

For convenience of presentation, in place of the projective labels $\mathbf{h}^{\pm} \!= (h_0^{\pm}, h_1^{\pm})$ we introduce the parameters
\begin{equation}\label{eq:bpm}
\gg_{\pm} := \pm \,(h_1^{\pm}/h_0^{\pm}) \in \mathbb{R} \cup \{+\infty\}\,,
\end{equation}
intending $\gg_{\pm} = +\infty$ if $h_0^{\pm} = 0$. Accordingly, the boundary conditions in Eq. \eqref{eq:Aperf} become
\begin{equation}\label{eq:bcb}
\mp \psi'(0^{\pm}) + \gg_{\pm}\, \psi(0^{\pm}) = 0\,,
\end{equation}
with the implication that $\psi(0^{\pm}) = 0$ if $\gg_{\pm} = +\infty$. Notice that we fixed the overall $\pm$ signs in the definition \eqref{eq:bpm} of $\gg_{\pm}$ so that both conditions in Eq. \eqref{eq:bcb} comply with the canonical form $(\partial_{n}\psi + \gg_{\pm} \psi)\big|_{x_1 \,=\, 0^{\pm}}\! = 0$, where $n$ denotes the unit outer normal. Of course, Neumann and Dirichlet boundary conditions are retrieved for $\gg_{\pm} = 0$ and $\gg_{\pm} = +\infty$, respectively. Let us also emphasize that, with our units of measure, the parameters $\gg_{\pm}$ are dimensionally equivalent to a mass.

For any $\gg_{+},\gg_{-} \in \mathbb{R} \cup \{+\infty\}$ the spectrum of the reduced operator $\mathcal{A}_1$ comprises an invariant purely absolutely continuous part and at most two isolated eigenvalues below the continuous threshold, depending on the sign of $\gg_{\pm}$. More precisely, we have
\begin{eqnarray*}
& \sigma(\mathcal{A}_1) = \sigma_{ac}(\mathcal{A}_1) \cup \sigma_p(\mathcal{A}_1)\,, \qquad
\sigma_{ac}(\mathcal{A}_1) = [m^2,+\infty)\,, \nonumber \\
& \sigma_{p}(\mathcal{A}_1) = \left\{\!\begin{array}{ll}
\displaystyle{ \varnothing }			& \displaystyle{ \mbox{for\, $\gg_{+},\gg_{-} \!\in\! [0,+\infty) \cup \{+\infty\}$}, } \vspace{0.1cm}\\
\displaystyle{ \big\{m^2 - \gg_{+}^2\big\} }	& \displaystyle{ \mbox{for\, $\gg_{+} \!\in\! (-\infty,0)$, $\gg_{-} \!\in\! [0,+\infty) \cup \{+\infty\}$}, } \vspace{0.1cm}\\
\displaystyle{ \big\{m^2 - \gg_{-}^2\big\} }	& \displaystyle{ \mbox{for\, $\gg_{+} \!\in\! [0,+\infty) \cup \{+\infty\}$, $\gg_{-} \!\in\! (-\infty,0)$}, } \vspace{0.1cm}\\
\displaystyle{ \big\{m^2 - \gg_{+}^2, m^2 - \gg_{-}^2\big\} }	& \displaystyle{ \mbox{for\, $\gg_{+},\gg_{-}\! \in\! (-\infty,0)$}. }
\end{array}\right. 
\end{eqnarray*}
This makes evident that the required positivity of $\mathcal{A}_1$ is ensured if and only if
\begin{equation}\label{eq:conbm}
\gg_{+},\gg_{-} \in (-m,+\infty) \cup \{+\infty\}\,, \qquad m > 0\,,
\end{equation}
two conditions which we assume to be fulfilled until the end of this section.

The heat kernel associated to the aforementioned reduced operator $\mathcal{A}_1$ is given by (see Appendix \ref{app:heatperf}; $\theta(\cdot)$ is the Heaviside step function)
\begin{align}\label{eq:heatrob}
e^{-\tau \mathcal{A}_1}(x_1,y_1) & = {e^{- m^2 \tau} \over \sqrt{4\pi \tau}} \left[
\theta(x_1)\, \theta(y_1) \left( 
e^{- {|x_1 - y_1|^2 \over 4\tau}}\! + e^{- {|x_1 + y_1|^2 \over 4\tau}} \!
- 2\gg_{+}\! \int_{0}^{\infty}\! dw\; e^{- \gg_{+} w \,- {(w + x_1 + y_1)^2 \over 4 \tau}}
\right)\right. \\
& \qquad \left. +\, \theta(-x_1)\, \theta(-y_1) \left(
e^{- {|x_1 - y_1|^2 \over 4\tau}}\! + e^{- {|x_1 + y_1|^2 \over 4\tau}}\!
- 2\gg_{-}\! \int_{0}^{\infty}\! dw\; e^{- \gg_{-} w \,- {(w - x_1 - y_1)^2 \over 4 \tau}}
\right) \right]. \nonumber 
\end{align}

Inserting this expression into Eq. \eqref{eq:fiuheat}, we obtain the following integral representation of the $\zeta$-regularized vacuum polarization:
\begin{align}\label{eq:fiuperf}
& \vacc \big(\hat{\phi}^{u}(t,\x)\big)^2 \vac 
= {\kappa^u \over 2\,(4\pi)^{d/2}\, \Gamma({u+1 \over 2})} \int_{0}^{\infty} d\tau\;\tau^{{u - d - 1 \over 2}}\, e^{- m^2 \tau} \;\times\\
& \qquad \times \!\left[
1 + e^{- {(x_1)^2 \over \tau}}\!
- 2\gg_{+}\,\theta(x_1)\!\int_{0}^{\infty}\!\! dw\; e^{- \gg_{+} w \,- {(w + 2 x_1)^2 \over 4 \tau}}\!
- 2\gg_{-}\,\theta(-x_1)\! \int_{0}^{\infty}\!\! dw\; e^{- \gg_{-} w \,- {(w - 2 x_1)^2 \over 4 \tau}}
\right]\,. \nonumber
\end{align}

In accordance with the general theory outlined in Section \ref{sec:gen}, it can be checked by direct inspection that the above representation \eqref{eq:fiuperf} makes sense for 
\begin{equation}\label{eq:Reu0}
u \in \mathbb{C} \quad \mbox{with} \quad \Re u > d - 1\,,
\end{equation}
a condition needed especially to ensure the convergence of the integral w.r.t. the variable $\tau$ for $\tau \to 0^{+}$. Besides, the expression of the right-hand side of Eq. \eqref{eq:fiuperf} is an analytic function of $u$ inside the semi-infinite complex strip identified by Eq. \eqref{eq:Reu0}.

In order to determine the meromorphic extensions of the map $u \mapsto \vacc \big(\hat{\phi}^{u}(t,\x)\big)^2 \vac $ to the whole complex plane, let us first make a brief digression to mention a pair of identities involving the Euler Gamma function $\Gamma$ and the modified Bessel function of second kind $K_{\nu}$. On one hand, we have \cite[Eq. 5.9.1]{NIST}
\begin{equation}\label{eq:idGamma}
\int_{0}^{\infty}\! d\tau\;\tau^{\nu - 1}\, e^{-m^2 \tau} = m^{-2\nu}\, \Gamma(\nu)\,,
\qquad \mbox{for all\; $m>0$, $\nu \in \mathbb{C}$ with $\Re \nu > 0$}\,;
\end{equation}
on the other hand, from \cite[Eq. 10.32.10]{NIST} we deduce
\begin{equation}\label{eq:idBesselK}
\int_{0}^{\infty}\! d\tau\;\tau^{\nu - 1}\, e^{-m^2 \tau - {p^2 \over \tau}}
= 2^{\nu + 1}\,p^{2\nu}\, \mathfrak{K}_{-\nu}(2 m p)\,,
\qquad \mbox{for all\; $m,p>0$,\, $\nu \in \mathbb{C}$}\,,
\end{equation}
where, for later convenience, we introduced the functions ($\nu \in \mathbb{C}$)
\begin{equation}\label{eq:KKdef}
\mathfrak{K}_{\nu} : (0,+\infty) \to \mathbb{C}\,, \qquad \mathfrak{K}_{\nu}(w) := w^{\nu}\,K_{\nu}(w)\,.
\end{equation}

Let us now return to Eq. \eqref{eq:fiuperf}. Firstly, notice that the integration order therein can be exchanged by Fubini's theorem, for any $u$ as in Eq. \eqref{eq:Reu0}. Then, using the previous identities \eqref{eq:idGamma} and \eqref{eq:idBesselK}, by a few additional manipulations we obtain
\begin{align*}
\vacc \big(\hat{\phi}^{u}(t,\x)\big)^2 \vac 
& = {\kappa^u \over 2\,(4\pi)^{d/2}\, \Gamma({u+1 \over 2})} \Bigg[
m^{d - 1 - u}\; \Gamma\left({u - d + 1 \over 2}\right) 
+ 2^{{u - d + 3 \over 2}}\,|x_1|^{u - d + 1}\, \mathfrak{K}_{{d - 1 - u \over 2}}\big(2 m |x_1|\big)
\\
& \qquad 
- 2^{{u - d + 5 \over 2}} \gg_{+}\,\theta(x_1)\!\int_{0}^{\infty}\!\! dw\; e^{- \gg_{+} w}\, (w/2 + x_1)^{u - d + 1}\, \mathfrak{K}_{{d - 1 - u \over 2}}\big(m (w/2 + x_1)\big)
\\
& \qquad
- 2^{{u - d + 5 \over 2}} \gg_{-}\,\theta(-x_1)\! \int_{0}^{\infty}\!\! dw\; e^{- \gg_{-} w}\, (w/2 - x_1)^{u - d + 1}\, \mathfrak{K}_{{d - 1 - u \over 2}}\big(m (w/2 - x_1)\big)
\Bigg]\,, \nonumber
\end{align*}
which, via the change of integration variable $w = 2 |x_1|\, v$, becomes
\begin{align}
\vacc \big(\hat{\phi}^{u}(t,\x)\big)^2 \vac 
& = {m^{d - 1}(\kappa/m)^u\, \Gamma({u - d + 1 \over 2}) \over 2^{d+1} \pi^{d/2}\, \Gamma({u+1 \over 2})} 
+ {2^{{u - 3d + 1 \over 2}}\, \big(\kappa |x_1|\big)^{\!u} \over \pi^{d/2}\, \Gamma({u+1 \over 2})\, |x_1|^{d - 1}} \, \mathfrak{K}_{{d - 1 - u \over 2}\!}\big(2 m |x_1|\big) \nonumber
\\
& \qquad
- {\theta(x_1)\; \gg_{+}\,2^{{u - 3d + 5 \over 2}} \big(\kappa |x_1|\big)^{\!u}  \over \pi^{d/2}\,\Gamma({u+1 \over 2})\,|x_1|^{d - 2}} \int_{0}^{\infty}\!\! dv\; {e^{- 2 \gg_{+} |x_1|\, v} \over (v + 1)^{d - 1 - u}}\, \mathfrak{K}_{{d - 1 - u \over 2}}\!\big(2 m |x_1| (v + 1)\big)\nonumber
\\
& \qquad
- {\theta(-x_1)\; \gg_{-}\, 2^{{u - 3d + 5 \over 2}} \big(\kappa |x_1|\big)^{\!u} \over \pi^{d/2}\,\Gamma({u+1 \over 2})\, |x_1|^{d - 2}} \int_{0}^{\infty}\!\! dv\; {e^{- 2 \gg_{-} |x_1|\, v} \over (v + 1)^{d - 1 - u}}\, \mathfrak{K}_{{d - 1 - u \over 2}}\big(2 m |x_1| ( v + 1)\big)\,. \label{eq:fiuAC}
\end{align}

Recall that the reciprocal of the Gamma function is analytic on the whole complex plane. Conversely, the Gamma function appearing in the numerator of the first term is a meromorphic function of u, with simple poles where its argument is equal to a non-positive integer, \textit{i.e.},
\begin{equation}\label{eq:poles}
u = d-1-2\ell\,, \qquad \mbox{with\; $\ell \in \{0,1,2,\dots\}$}\,.
\end{equation}
On the other hand, from basic features of the modified Bessel function $K_{\nu}$ 
we infer that the function $\mathfrak{K}_{\nu}$ introduced in Eq. \eqref{eq:KKdef} fulfils the following: for any fixed $w > 0$, the map $\nu \mapsto \mathfrak{K}_{\nu}(w)$ is analytic on the whole complex plane \cite[\S 10.25(ii)]{NIST}; for any fixed $\nu \in \mathbb{C}$, the map $w \in (0,+\infty) \mapsto \mathfrak{K}_{\nu}(w)$ is analytic, continuous up to $w = 0$ and decaying with exponential speed for $w \to +\infty$ \cite[\S 10.31 and Eqs. 10.25.2, 10.27.4, 10.40.2]{NIST}.\footnote{Consider the integrals appearing in Eq. \eqref{eq:fiuAC}. Since the integrand functions therein are continuous at $w = 0$, the lower extreme of integration is never problematic. On the other hand, from \cite[10.40.2]{NIST} we infer
$$ e^{- 2 \gg_{\pm} |x_1|\, v}\, (v + 1)^{u - d + 1} \mathfrak{K}_{{d - 1 - u \over 2}}\big(2 m |x_1| ( v + 1)\big)
= \sqrt{\pi \over 2}\; e^{- 2 m |x_1|} \big(2 m |x_1|\big)^{{d - 2 - u \over 2}}\;
e^{- 2 (\gg_{\pm} + m)\, |x_1|\, v}\, v^{{u - d \over 2}} \Big(1 + o(1)\Big)
\quad \mbox{for\; $w \to +\infty$}\,, $$
which shows that the condition $\gg_{\pm} > - m$ established in Eq. \eqref{eq:conbm} is in fact indispensable to grant the convergence of the said integrals.}

In light of the above considerations, Eq. \eqref{eq:fiuAC} does in fact provide the meromorphic extension of the $\zeta$-regularized vacuum polarization $\vacc \big(\hat{\phi}^{u}(t,\x)\big)^2 \vac$ to the whole complex plane, with isolated simple pole singularities at the points indicated in Eq. \eqref{eq:poles}. We can then proceed to compute the renormalized vacuum polarization, implementing the general prescription \eqref{eq:fi2ren}. In this regard, special attention must be paid to the first term on the right-hand side of Eq. \eqref{eq:fiuAC}, since it presents a pole at $u = 0$ when the space dimension $d$ is odd. More precisely, using some basic properties of the Gamma function \cite[\S 5]{NIST} and indicating with $H_{\ell} := \sum_{j = 1}^{\ell} {1 \over j}$ the $\ell$-th harmonic number for $\ell \in \{0,1,2, \dots\}$ ($H_{0} \equiv 0$ by convention), we deduce
\begin{equation*}
{(\kappa/m)^u\, \Gamma({u - d + 1 \over 2}) \over \Gamma({u+1 \over 2})} = \left\{\!\!\begin{array}{ll}
\displaystyle{{(-1)^{{d \over 2}}\,\sqrt{\pi} \over \Gamma({d + 1 \over 2})} + \mathcal{O}(u)}	&	 \displaystyle{\mbox{for $d$ even}\,,} \vspace{0.1cm}\\
\displaystyle{{(-1)^{{d-1 \over 2}} \over \sqrt{\pi}\; \Gamma({d+1 \over 2})} \left[{2 \over u} + H_{{d-1 \over 2}} + 2\log \left({2\kappa \over m}\right) \right] + \mathcal{O}(u)}	&	 \displaystyle{\mbox{for $d$ odd}\,.}
\end{array}\right.
\end{equation*}

Noting that all other terms in Eq. \eqref{eq:fiuAC} are regular at $u = 0$, from \eqref{eq:fi2ren} we obtain
\begin{equation}\label{eq:firen}
\vacc \hat{\phi}^{2}(t,\x) \vac_{ren} = \vacc \hat{\phi}^{2} \vac_{ren}^{(free)} + \vacc \hat{\phi}^{2}(x_1) \vac_{ren}^{(plane)}\,;
\end{equation}
\begin{equation}\label{eq:firenm}
\vacc \hat{\phi}^{2} \vac_{ren}^{(free)} = \left\{\!\!\begin{array}{ll}
\displaystyle{ {(-1)^{{d \over 2}}\,\pi\;m^{d - 1} \over (4 \pi)^{{d + 1 \over 2}}\; \Gamma({d + 1 \over 2})} }	&	 \displaystyle{\mbox{for $d$ even}\,,} \vspace{0.1cm} \\
\displaystyle{ {(-1)^{{d-1 \over 2}}\; m^{d - 1} \over (4 \pi)^{{d + 1 \over 2}}\; \Gamma({d+1 \over 2})} \left[H_{{d-1 \over 2}} + 2\log \left({2\kappa \over m}\right) \right]}	&	 \displaystyle{\mbox{for $d$ odd}\,;}
\end{array}\right.
\end{equation}
\begin{align}\label{eq:firenpi}
\vacc \hat{\phi}^{2}(x_1) \vac_{ren}^{(plane)} 
& := {1 \over 2^{{3d - 1 \over 2}} \pi^{{d+1 \over 2}}\, |x_1|^{d - 1}}\, \Bigg[
\mathfrak{K}_{{d - 1 \over 2}\!}\big(2 m |x_1|\big) 
\\
& \qquad
- \theta(x_1)\; 4 \gg_{+} |x_1| \int_{0}^{\infty}\!\! dv\; {e^{- 2 \gg_{+} |x_1|\, v} \over (v + 1)^{ d - 1}}\; \mathfrak{K}_{{d - 1 \over 2}} \big(2 m |x_1| (v + 1)\big)\nonumber
\\
& \qquad
- \theta(-x_1)\; 4 \gg_{-}  |x_1| \int_{0}^{\infty}\!\! dv\; {e^{- 2 \gg_{-} |x_1|\, v} \over (v + 1)^{ d - 1}}\; \mathfrak{K}_{{d - 1 \over 2}}\big(2 m |x_1| ( v + 1)\big)\Bigg]\,. \nonumber
\end{align}

It is worth remarking that $\vacc \hat{\phi}^{2}\vac_{ren}^{(free)}$ is in fact a constant which depends solely on the mass $m$ of the field and, possibly, on the renormalization mass parameter $\kappa$ (if the space dimension $d$ is odd). In particular, it does not depend on the coordinate $x_1$, namely, the distance from the hyperplane $\pi$, nor on the parameters $\gg_{\pm}$ defining the boundary conditions on $\pi$. For these reasons it is natural to regard $\vacc \hat{\phi}^{2} \vac_{ren}^{(free)}$ as a pure free-theory contribution (which explains the choice of the superscript). In contrast, $\vacc \hat{\phi}^{2}(x_1) \vac_{ren}^{(plane)}$ is a contribution which truly accounts for the presence of the hyperplane and for the boundary conditions on it.

Owing to the above considerations, one might be tempted to discard the free-theory term $\vacc \hat{\phi}^{2}\vac_{ren}^{(free)}$ and regard $\vacc \hat{\phi}^{2}(x_1) \vac_{ren}^{(plane)}$ as the only physically relevant contribution to the vacuum polarization. Despite being tenable, this standpoint actually suffers from a drawback. Indeed, let us anticipate that $\vacc \hat{\phi}^{2}\vac_{ren}^{(free)}$ plays a key role in the cancellation of some infrared divergences which would otherwise affect the massless theory in space dimension $d = 1$ (see the subsequent paragraph  \ref{subsubsec:m0d1}). Therefore, we reject the standpoint sketched above and regard the sum $\vacc \hat{\phi}^{2}(t,\x) \vac_{ren}$ defined in Eq. \eqref{eq:firen} as the true physically sensible observable.

\subsection{Neumann and Dirichlet conditions}

We already mentioned in the comments below Eq. \eqref{eq:bcb} that Neumann and Dirichlet boundary conditions correspond to $\gg_{\pm} = 0$ and $\gg_{\pm} = +\infty$, respectively. Of course, the free-theory contribution $\vacc \hat{\phi}^{2} \vac_{ren}^{(free)}$ remains unchanged in both cases, so let us focus on the term $\vacc \hat{\phi}^{2}(x_1) \vac_{ren}^{(plane)}$.

In the case of Neumann conditions where $\gg_{\pm} = 0$, it appears that the expressions in the second and third line of Eq. \eqref{eq:firenpi} vanish identically. Regarding the case of Dirichlet conditions, the limits $\gg_{\pm} \to +\infty$ can be easily computed as follows, making the change of integration variable $z = 2 \gg_{+} |x_1|\, v$ and using the dominated convergence theorem:
\begin{align*}
& \lim_{\gg_{\pm} \to +\infty} \Bigg[4 \gg_{\pm} |x_1| \int_{0}^{\infty}\!\! dv\; {e^{- 2 \gg_{\pm} |x_1|\, v} \over (v + 1)^{ d - 1}}\; \mathfrak{K}_{{d - 1 \over 2}}\big(2 m |x_1| (v + 1)\big) \Bigg]\\
& = \lim_{\gg_{\pm} \to +\infty} \Bigg[2 \int_{0}^{\infty}\!\! dz\; {e^{- z} \over \big({z \over 2 \gg_{+} |x_1|} + 1\big)^{ d - 1}}\; \mathfrak{K}_{{d - 1 \over 2}}\!\left(2 m |x_1| \Big({z \over 2 \gg_{+} |x_1|} + 1\Big)\right) \Bigg] \\
& = 2\, \mathfrak{K}_{{d - 1 \over 2}}\!\big(2 m |x_1|\big) \int_{0}^{\infty}\!\! dz\; e^{- z}
= 2\, \mathfrak{K}_{{d - 1 \over 2}}\!\big(2 m |x_1|\big)\,.
\end{align*}

Summarizing, Eq. \eqref{eq:firenpi} reduces to
\begin{equation}\label{eq:firenpiDN}
\vacc \hat{\phi}^{2}(x_1) \vac_{ren}^{(plane)} 
= \pm\,{\mathfrak{K}_{{d - 1 \over 2}\!}\big(2 m |x_1|\big) \over 2^{{3d - 1 \over 2}} \pi^{{d+1 \over 2}}\, |x_1|^{d - 1}}\,,
\end{equation}
for Neumann (+) and Dirichlet (-) boundary conditions, respectively.

\subsection{Asymptotics for $x_1 \to 0^{\pm}$ and $x_1 \to \pm \infty$}\label{subsec:asy}

Hereafter we investigate the behaviour of the renormalized vacuum polarization $\vacc \hat{\phi}^{2}(t,\x) \vac_{ren}$ close to the hyperplane $\pi$ and far way from it. For brevity we only present the leading order asymptotics, although a refinement of the arguments outlined below would actually permit to derive asymptotic expansions at any order.

Before proceeding, let us stress once more that $\vacc \hat{\phi}^{2} \vac_{ren}^{(free)}$ does not depend on the coordinate $x_1$; thus, it is sufficient to analyse the term $\vacc \hat{\phi}^{2}(x_1) \vac_{ren}^{(plane)}$ (see Eq. \eqref{eq:firenpi}).

\subsubsection{The limit $x_1 \to 0^{\pm}$}\label{par:x10}

Let us first notice that the functions $\mathfrak{K}_{\nu}$ defined in Eq. \eqref{eq:KKdef} have the following asymptotic expansions, which can be easily derived from \cite{NIST}[Eqs. 10.31.1 and 10.30.2] (here and in the sequel $\gamma_{EM} = 0.57721\dots$ indicates the Euler-Mascheroni constant):
\begin{eqnarray}
& \mathfrak{K}_{0}(w) = - \log(w/2) + \gamma_{EM} + \mathcal{O}\big(w^2 \log w\big)\,, \qquad \mbox{for\; $w \to 0^{+}$}\,; \label{eq:K0asy} \\
& \mathfrak{K}_{\nu}(w) = 2^{\nu-1}\,\Gamma(\nu) + \mathcal{O}\big(w^{\min\{2+\nu,\,2\nu\}} \log w\big) \,, \qquad \mbox{for\, $w \to 0^{+}$ and\, $\nu > 0$}\,. \label{eq:Knuasy}
\end{eqnarray}

Next, consider the integrals appearing in the second and third lines of Eq. \eqref{eq:firenpi}. For any finite $\gg_{\pm} \in (-m,+\infty)$ (cf. Eq. \eqref{eq:conbm}), via the change of variable $z = 2m |x_1|\,(v+1)$ we get
\begin{align*}
& 4 \gg_{\pm} |x_1| \int_{0}^{\infty}\!\! dv\; {e^{- 2 \gg_{\pm} |x_1|\, v} \over (v + 1)^{ d - 1}}\; \mathfrak{K}_{{d - 1 \over 2}} \big(2 m |x_1| (v + 1)\big) 
= {2 \gg_{\pm} \over m}\,\big(2m |x_1|\big)^{d - 1}\, e^{2 \gg_{\pm} |x_1|} \int_{2 m |x_1|}^{\infty}\! dz\; {e^{- {\gg_{\pm} \over m}\, z} \over z^{d - 1}}\; \mathfrak{K}_{{d - 1 \over 2}}(z)
\end{align*}

Writing $\int_{2 m |x_1|}^{\infty} = \int_{2 m |x_1|}^{z_0} + \int_{z_0}^{\infty}$ for some $z_0 > 0$ fixed arbitrarily and replacing the integrand inside $\int_{2 m |x_1|}^{z_0}$ with its Taylor expansion at $z = 0$ (recall, especially, Eqs. \eqref{eq:K0asy} and \eqref{eq:Knuasy}), by a few additional computations we deduce, for $|x_1| \to 0^{+}$,
\begin{align*}
& 4 \gg_{\pm} |x_1| \int_{0}^{\infty}\!\! dv\; {e^{- 2 \gg_{\pm} |x_1|\, v} \over (v + 1)^{ d - 1}}\; \mathfrak{K}_{{d - 1 \over 2}} \big(2 m |x_1| (v + 1)\big)
= \left\{\! \begin{array}{ll}
\displaystyle{ \mathcal{O}(1) }	&	\displaystyle{\mbox{for\; $d = 1$}\,,} \vspace{0.1cm} \\
\displaystyle{ \mathcal{O}\big( |x_1|\, \log |x_1|\big) }	&	\displaystyle{\mbox{for\; $d = 2$}\,,} \vspace{0.1cm} \\
\displaystyle{ \mathcal{O}\big(|x_1|\big) }	&	\displaystyle{\mbox{for\; $d \geqslant 3$}\,.} 
\end{array} \right.
\end{align*}

Summing up, the above arguments allow us to infer that, in the limit $x_1 \to 0^{\pm}$,
\begin{align}\label{eq:firenx0}
& \vacc \hat{\phi}^{2}(x_1) \vac_{ren}^{(plane)}
= \left\{\! \begin{array}{ll}
\displaystyle{ -\,{1 \over 2 \pi}\,\log\big(m |x_1|\big) + \mathcal{O}(1) }	&	\displaystyle{\mbox{for\; $d = 1$}\,,} \vspace{0.1cm} \\
\displaystyle{ {1 \over 8 \pi\, |x_1|} + \mathcal{O}\big( \log |x_1|\big) }	&	\displaystyle{\mbox{for\; $d = 2$}\,,} \vspace{0.1cm} \\
\displaystyle{ {\Gamma({d - 1 \over 2}) \over (4 \pi)^{{d+1 \over 2}}\, |x_1|^{d - 1}} \Big[1 + \mathcal{O}\big(|x_1|\big)\Big] }	&	\displaystyle{\mbox{for\; $d \geqslant 3$}\,.} 
\end{array} \right.
\end{align}

It is remarkable that the above leading order expansions do not depend on the parameters $\gg_{\pm}$, describing the boundary conditions. In particular, the same results remain valid for Neumann conditions, corresponding to $\gg_{\pm} = 0$. On the contrary, a separate analysis is required for Dirichlet conditions, which is formally recovered for $\gg_{\pm} \to + \infty$ (a limit which clearly does not commute with $x_1 \to 0^{\pm}$); in this case, starting from Eq. \eqref{eq:firenpiDN} and using again Eqs. \eqref{eq:K0asy} \eqref{eq:Knuasy} one can derive asymptotic expansions which coincide with those reported in Eq. \eqref{eq:firenx0}, except for the opposite overall sign.

\subsubsection{The limit $x_1 \to \pm\infty$}

It is a well known fact that local observables of Casimir type for massive fields are typically suppressed with exponential rate in the regime of large distances from the boundaries. In the sequel we provide quantitative estimates for $\vacc \hat{\phi}^{2}(x_1) \vac_{ren}^{(plane)}$, confirming this general expectation.
To this purpose, let us first point out that the functions $\mathfrak{K}_{\nu}$ fulfil (see Eq. \eqref{eq:KKdef} and \cite{NIST}[Eq. 10.40.2])
\begin{equation}\label{eq:Kinf}
\mathfrak{K}_{\nu}(w) = \sqrt{{\pi \over 2}}\;e^{-w}\,w^{\nu-1/2}\, \Big(1 + \mathcal{O}(1/w) \Big) \,, \qquad \mbox{for\, $w \to +\infty$\, and\, $\nu \geqslant 0$}\,.
\end{equation}

Consider now the integral expressions in Eq. \eqref{eq:firenpi}. Using the above relation and making the change of variable $z = |x_1|\, v$, for $|x_1| \to +\infty$ we deduce
\begin{align*}
& 4 \gg_{\pm} |x_1| \int_{0}^{\infty}\!\! dv\; {e^{- 2 \gg_{\pm} |x_1|\, v} \over (v + 1)^{ d - 1}}\; 
\mathfrak{K}_{{d - 1 \over 2}} \big(2 m |x_1| (v + 1)\big) \\
& = 2 \sqrt{2 \pi}\; \gg_{\pm} \;e^{- 2 m |x_1|}\,\big(2 m |x_1|\big)^{{d - 2 \over 2}} \int_{0}^{\infty}\!\! dz\; e^{- 2 (\gg_{\pm} + m)\,z}\, \Big(1 + \mathcal{O}\big(z/|x_1|\big) \Big) \\
& = {\sqrt{2 \pi}\; \gg_{\pm} \over \gg_{\pm} + m} \;e^{- 2 m |x_1|}\,\big(2 m |x_1|\big)^{{d - 2 \over 2}} \Big(1 + \mathcal{O}\big(1/|x_1|\big) \Big)\,.
\end{align*}

In view of the above results, from Eq. \eqref{eq:firenpi} we infer
\begin{equation}\label{eq:fireninf}
\vacc \hat{\phi}^{2}(x_1) \vac_{ren}^{(plane)}\!
 = {m^{{d-2 \over 2}} \over 2 (4 \pi)^{{d/2}}} \left({m \!-\! \gg_{\pm} \over m \!+\! \gg_{\pm}}\right) {e^{- 2 m |x_1|} \over |x_1|^{d/2}}\, \left[1 + \mathcal{O}\left({1 \over |x_1|} \right) \right] 
\qquad \mbox{for\, $x_1 \!\to\! \pm \infty$} \,. 
\end{equation}

The case of Dirichlet boundary conditions can be alternatively addressed taking the limit $\gg_{\pm} \to +\infty$ in Eq. \eqref{eq:fireninf}, or starting from Eq. \eqref{eq:firenpiDN} and using again Eq. \eqref{eq:Kinf}:
\begin{equation*}
\vacc \hat{\phi}^{2}(x_1) \vac_{ren}^{(plane)}\!
 = -\,{m^{{d-2 \over 2}} \over 2 (4 \pi)^{{d/2}}}\; {e^{- 2 m |x_1|} \over |x_1|^{d/2}}\, \left[1 + \mathcal{O}\left({1 \over |x_1|} \right) \right] 
\qquad \mbox{for $x_1 \!\to\! \pm \infty$} \,.
\end{equation*}

As a final remark, let us highlight that in any case $\vacc \hat{\phi}^{2}(t,\x) \vac_{ren}$ approaches the constant free-theory value $\vacc \hat{\phi}^{2} \vac_{ren}^{(free)}$ with exponential speed.

\subsection{Vacuum polarization for a massless field}\label{subsec:m0}

Let us now address the case of a massless field, fulfilling generic boundary conditions of the form \eqref{eq:bcb}. In this context the hypothesis \eqref{eq:conbm} entails
\begin{equation*}
\gg_{+},\gg_{-} \in [0,+\infty) \cup \{+\infty\}\,,
\end{equation*}

and under this condition we can implement the general arguments reported in Section \ref{sec:gen}. Especially, let us recall that the renormalized vacuum polarization for a massless field is obtained as the zero-mass limit of the analogous quantity for a massive field, see Eq. \eqref{eq:fi2renm0}. In the sequel we discuss separately the cases with space dimension $d = 1$ and $d \geqslant 2$, for both technical and physical reasons.

\subsubsection{Space dimension $d = 1$}\label{subsubsec:m0d1}
This case deserves a separate analysis, due to the emergence of some delicate infrared features. As a matter of fact, both $\vacc \hat{\phi}^{2} \vac_{ren}^{(free)}$ and $\vacc \hat{\phi}^{2}(x_1) \vac_{ren}^{(plane)}$ diverge in the limit $m \to 0^{+}$; however, their sum $\vacc \hat{\phi}^{2}(t,\x) \vac_{ren}$ remains finite, except when the boundary conditions are of Neumann type.

To account for the above claims, let us firstly notice that Eq. \eqref{eq:firenm} yields (for $d = 1$)
\begin{equation}\label{eq:firenm1d}
\vacc \hat{\phi}^{2}\vac_{ren}^{(free)} = {1 \over 2 \pi}\, \log \left({2\kappa \over m}\right) ,
\end{equation}

which is patently divergent in the limit $m \to 0^{+}$. 

Now consider the term $\vacc \hat{\phi}^{2}(x_1) \vac_{ren}^{(plane)}$. For $\gg_{+} = \gg_{-} = 0$,\footnote{Similar results can be derived also if only one of $\gg_{+}$ and $\gg_{-}$ is equal to zero.} namely in the case of Neumann conditions, from Eqs. \eqref{eq:firenpiDN} and \eqref{eq:K0asy} we readily infer (for fixed $x_1 \neq 0$)
\begin{equation*}
\vacc \hat{\phi}^{2}(x_1) \vac_{ren}^{(plane)} 
= - \,{1 \over 2 \pi}\,\log\big(m |x_1|\big) + \mathcal{O}(1)\,, \qquad \mbox{for\; $m \to 0^{+}$}\,,
\end{equation*}
which, together with Eqs. \eqref{eq:firen} and \eqref{eq:firenm1d}, implies in turn
\begin{equation*}
\lim_{m \to 0^{+}} \vacc \hat{\phi}^{2}(t,\x) \vac_{ren} = 
\lim_{m \to 0^{+}} \left[{1 \over 2 \pi}\, \log \left({\kappa \over m^2 |x_1|}\right) + \mathcal{O}(1)\right] = + \infty\,.
\end{equation*}

This is nothing but an unavoidable manifestation of the infrared divergences which typically affect massless theories in low space dimension. Taking notice of this fact, in the remainder of this subsection we restrict the attention to
\begin{equation*}
\gg_{+},\gg_{-} \in (0,+\infty)\,.
\end{equation*}
With this requirement, using Eq. \eqref{eq:K0asy} and some known integral identities for the incomplete Gamma function $\Gamma(a,z)$,\footnote{More precisely, integrating by parts, making the change of integration variable $z = 2 \gg_{\pm} |x_1|\, (v+1)$ and recalling \cite[Eq. 8.2.2]{NIST} one gets
$$ 2 \gg_{\pm} |x_1| \int_{0}^{\infty}\!\! dv\; e^{- 2 \gg_{\pm} |x_1|\, v}\,\log(v+1)
= - \Big(e^{- 2 \gg_{\pm} |x_1|\, v}\,\log(v\!+\!1)\Big)_{0}^{+ \infty}
\!+ \int_{0}^{\infty}\!\! dv\; {e^{- 2 \gg_{\pm} |x_1|\, v} \over v+1}
= e^{2 \gg_{\pm} |x_1|} \int_{2 \gg_{\pm} |x_1|}^{\infty}\!\! dz\; {e^{- z}  \over z}
= e^{2 \gg_{\pm} |x_1|}\, \Gamma\big(0,2 \gg_{\pm} |x_1|\big)\,. $$
}
from Eq. \eqref{eq:firenpi} we deduce the following for $m \to 0^{+}$:
\begin{align*}
\vacc \hat{\phi}^{2}(x_1) \vac_{ren}^{(plane)} 
& = {1 \over 2 \pi}\, \Bigg[ - \log\big(m |x_1|\big) + \gamma_{EM}
\nonumber  \\
& \qquad\qquad - \theta(x_1)\; 4 \gg_{+} |x_1| \int_{0}^{\infty}\!\! dv\; e^{- 2 \gg_{+} |x_1|\, v}\; 
\Big(- \log\big(m |x_1|\big) + \gamma_{EM} - \log(v + 1)\Big)
\nonumber \\
& \qquad\qquad - \theta(-x_1)\; 4 \gg_{-}  |x_1| \int_{0}^{\infty}\!\! dv\; e^{- 2 \gg_{-} |x_1|\, v}\; 
\Big(- \log\big(m |x_1|\big) + \gamma_{EM} - \log(v + 1) \Big) \Bigg] 
\nonumber \\
& \qquad\qquad + \mathcal{O}\Big(\big(m |x_1|\big)^2 \log \big(m |x_1|\big)\Big)
\\
& = {1 \over 2 \pi}\, \Big[
\log\big(m |x_1|\big) - \gamma_{EM}
+ 2\, \theta(x_1)\; e^{2 \gg_{+} |x_1|}\, \Gamma\big(0,2 \gg_{+} |x_1|\big)
\nonumber  \\
& \qquad\qquad
+ 2\, \theta(-x_1)\; e^{2 \gg_{-} |x_1|}\, \Gamma\big(0,2 \gg_{-} |x_1|\big)\Big]
+ \mathcal{O}\Big(\big(m |x_1|\big)^2 \log \big(m |x_1|\big)\Big)\,. 
\nonumber
\end{align*}
From here and from Eqs. \eqref{eq:fi2renm0} and \eqref{eq:firenm1d}, we finally obtain
\begin{align}
& \vacc \hat{\phi}^{2}(t,\x) \vac_{ren}^{(massless)} 
= \lim_{m \to 0^{+}} \Big[ \vacc \hat{\phi}^{2} \vac_{ren}^{(free)} + \vacc \hat{\phi}^{2}(x_1) \vac_{ren}^{(plane)} \Big] \nonumber \\
& = {1 \over 2 \pi}\, \Big[
\log \big(2\kappa |x_1|\big) - \gamma_{EM}
+ 2\, \theta(x_1)\; e^{2 \gg_{+} |x_1|}\, \Gamma\big(0,2 \gg_{+} |x_1|\big)
+ 2\, \theta(-x_1)\; e^{2 \gg_{-} |x_1|}\, \Gamma\big(0,2 \gg_{-} |x_1|\big)\Big]\,.
\label{eq:firend1m0}
\end{align}

The case of Dirichlet boundary conditions is retrieved taking the limit $\gg_{\pm} \to +\infty$ and noting that the incomplete Gamma function fulfils $\lim_{w \to +\infty} e^{w}\, \Gamma(0,w) = 0$ (see \cite[Eq. 8.11.2]{NIST}), which gives
\begin{equation*}
\vacc \hat{\phi}^{2}(t,\x) \vac_{ren}^{(massless)} = {1 \over 2 \pi}\, \Big[\log \big(2\kappa |x_1|\big) - \gamma_{EM} \Big]\,. 
\end{equation*}

For any $\gg_{+},\gg_{-} \in (0,+\infty)$, the asymptotic behaviour of $\vacc \hat{\phi}^{2}(t,\x) \vac_{ren}^{(massless)}$ for small and large distances from the point $\pi \equiv \{x_1 = 0\}$ can be easily derived from the explicit expression \eqref{eq:firend1m0}, using the known series expansions for the incomplete Gamma function (see \cite[Eqs. 8.7.6 and 8.11.2]{NIST}). More precisely, to leading order we have
\begin{equation*}
\vacc \hat{\phi}^{2}(t,\x) \vac_{ren}^{(massless)} = \left\{\!\!\begin{array}{ll}
\displaystyle{ -\, {1 \over 2 \pi} \log \big(\kappa |x_1|\big) + \mathcal{O}(1) }	&	\displaystyle{\mbox{for\, $x_1 \to 0^{\pm}$},} \vspace{0.1cm} \\
\displaystyle{ {1 \over 2 \pi} \log\big(\kappa |x_1|\big)  + \mathcal{O}(1) }	&	\displaystyle{\mbox{for\, $x_1 \to \pm \infty$}\,.}
\end{array}
\right.
\end{equation*}

\subsubsection{Space dimension $d \geqslant 2$}\label{subsubsec:d2perfm0}
In this case it can be easily checked that the free-theory contribution $\vacc \hat{\phi}^{2} \vac_{ren}^{(free)}$ vanishes in the limit $m \to 0^{+}$ (see Eq. \eqref{eq:firenm}). Bearing this in mind, let us focus on the term $\vacc \hat{\phi}^{2}(x_1) \vac_{ren}^{(plane)}$. Recalling the asymptotic relation \eqref{eq:Knuasy} for $\mathfrak{K}_\nu$, by dominated convergence from Eq. \eqref{eq:firenpi} we infer
\begin{align*}
\lim_{m \to 0^{+}} \vacc \hat{\phi}^{2}(x_1) \vac_{ren}^{(plane)} 
& = {\Gamma({d - 1 \over 2}) \over (4 \pi)^{{d+1 \over 2}} |x_1|^{d - 1}} \left[
1 - \theta(x_1)\; 4 \gg_{+} |x_1| \int_{0}^{\infty}\!\! dv\; {e^{- 2 \gg_{+} |x_1|\, v} \over (v + 1)^{ d - 1}} \right.
\nonumber \\
& \hspace{3.5cm} \left.
-\, \theta(-x_1)\; 4 \gg_{-}  |x_1| \int_{0}^{\infty}\!\! dv\; {e^{- 2 \gg_{-} |x_1|\, v} \over (v + 1)^{d - 1}}\right] . \nonumber
\end{align*}

To say more, via the change of variable $z = 2 \gg_{\pm} |x_1| (v+1)$, the above integrals can be expressed in terms of incomplete Gamma functions $\Gamma(a,z)$ (see \cite[Eq. 8.2.2]{NIST}). 
Summing up, we ultimately obtain
\begin{align}
& \vacc \hat{\phi}^{2}(t,\x) \vac_{ren}^{(massless)}\! 
= \lim_{m \to 0^{+}} \vacc \hat{\phi}^{2}(x_1) \vac_{ren}^{(plane)} \nonumber \\
& = {\Gamma({d - 1 \over 2}) \over (4 \pi)^{{d+1 \over 2}} |x_1|^{d - 1}} \left[
1 - 2\,\theta(x_1)\, (2 \gg_{+} |x_1|)^{d-1} e^{2 \gg_{+} |x_1|}\, \Gamma\big(2\!-\!d, 2\gg_{+} |x_1|\big) \right.
\nonumber \\
& \hspace{3.5cm} \left.
-\, 2\,\theta(-x_1)\, (2 \gg_{-} |x_1|)^{d-1} e^{2 \gg_{-} |x_1|}\, \Gamma\big(2-d, 2\gg_{-} |x_1|\big) \right] . \label{eq:firenm0}
\end{align}

The renormalized vacuum polarization for a massless field subject to Neumann or Dirichlet boundary conditions can be deduced from the above result evaluating the limits $\gg_{\pm} \to 0^{+}$ or $\gg_{\pm} \to +\infty$, respectively. To be more precise, taking into account that $\lim_{w \to 0^{+}} w^{1-a} e^{w} \Gamma(a,w) = 0$ and $\lim_{w \to +\infty} w^{1-a} e^{w} \Gamma(a,w) = 1$ (see \cite[Eqs. 8.7.6 and 8.11.2]{NIST}), for Neumann (+) and Dirichlet (-) conditions we get
\begin{equation*}
\vacc \hat{\phi}^{2}(t,\x) \vac_{ren}^{(massless)} = \pm \,{\Gamma({d - 1 \over 2}) \over (4 \pi)^{{d+1 \over 2}} |x_1|^{d - 1}} \;.
\end{equation*}
The same result can be alternatively derived using \eqref{eq:Knuasy} to compute the limit $m \to 0^{+}$ of Eq. \eqref{eq:firenpiDN}.
Also in this case, for any $\gg_{+},\gg_{-} \in (0,+\infty)$ the behaviour of $\vacc \hat{\phi}^{2}(t,\x) \vac_{ren}^{(massless)}$ for $x_1 \to 0^{\pm}$ and $x_1 \to \pm \infty$ can be inferred from Eq. \eqref{eq:firenm0} using the corresponding expansions for the incomplete Gamma function (see \cite[Eqs. 8.7.6 and 8.11.2]{NIST}). To leading order, we have
\begin{equation*}
\vacc \hat{\phi}^{2}(t,\x) \vac_{ren}^{(massless)} = \left\{\!\!\begin{array}{ll}
\displaystyle{ {1 \over 8 \pi\, |x_1|} + \mathcal{O}\big(\log|x_1| \big) }	&	\displaystyle{\mbox{for\, $d = 2$,\, $x_1 \to 0^{\pm}$},} \vspace{0.1cm} \\
\displaystyle{ {\Gamma({d - 1 \over 2}) \over (4 \pi)^{{d+1 \over 2}} |x_1|^{d - 1}} \Big[1 + \mathcal{O}\big(|x_1| \big) \Big] }	&	\displaystyle{\mbox{for\, $d \geqslant 3$,\, $x_1 \to 0^{\pm}$},} \vspace{0.1cm} \\
\displaystyle{ -\, {\Gamma({d - 1 \over 2}) \over (4 \pi)^{{d+1 \over 2}} |x_1|^{d - 1}} \Big[1 + \mathcal{O}\big(1/|x_1| \big) \Big] }	&	\displaystyle{\mbox{for\, $x_1 \to \pm \infty$}\,.}
\end{array}
\right.
\end{equation*}

\section{Semitransparent plane}\label{sec:semitr}

Let us now examine configurations where the hyperplane $\pi$ can be regarded as a semitransparent surface. In this connection, recalling the general arguments of Section \ref{sec:gen} and referring again to \cite[Thm. 3.2.3]{AK99}, we consider the family of reduced operators labelled as follows by the elements of the unitary group $U(2)$:
\begin{eqnarray}
& \mbox{dom}(\mathcal{A}_1) := \left\{\psi \in H^2(\mathbb{R} \!\setminus\! \{0\})\,\left|\, 
\left(\! \begin{array}{c} \displaystyle{\psi(0^{+})} \\ \displaystyle{\psi'(0^{+})}  \end{array} \!\right) = \omega \left(\! \begin{array}{cc} \aa & \bb \\ \cc & \di \end{array} \!\right) \left(\! \begin{array}{c} \displaystyle{\psi(0^{-})} \\ \displaystyle{\psi'(0^{-})} \end{array} \! \right)\right.\right\} \,, \nonumber\\
& \mathcal{A}_1\, \psi = (-\,\partial_{x_1 x_1}\! + m^2)\, \psi \quad \mbox{in\; $\mathbb{R} \!\setminus\! \{0\}$}\,,\label{eq:Asemi}
\end{eqnarray}
where
\begin{equation}
\mbox{$\omega \in \mathbb{C}$\; with\; $|\omega| = 1$} \qquad \mbox{and} \qquad
\mbox{$\aa,\bb,\cc,\di \in \mathbb{R}$\; with\; $\aa \di - \bb \cc = 1$}\,. \label{eq:omabcd}
\end{equation}

Two distinguished one-parameter subfamilies are respectively obtained for either $\bb = 0$, $\cc \in \mathbb{R}$, $\omega = \aa = \di = 1$ or $\bb \in \mathbb{R}$, $\cc = 0$, $\omega = \aa = \di = 1$. These formally correspond to reduced operators of the form $\mathcal{A}_1 = -\, \partial_{x_1 x_1} + m^2 + \cc\,\delta$ or $\mathcal{A}_1 = - \,\partial_{x_1 x_1} + m^2 + \bb\,\delta'$, containing the well-known distributional delta and delta-prime potentials. The other admissible choices of parameters formally correspond to mixtures of delta and delta-prime potentials concentrated at $x_1 = 0$ (see \cite{Se86} and \cite[\S 3.2.4]{AK99}). Let us further remark that for $\bb = \cc = 0$ and $\omega = \aa = \di = 1$ the reduced operator $\mathcal{A}_1$ is just the free Laplacian on the line; this case corresponds to a configuration where the quantum field does not interact with the plane $\pi$.

For any choice of the parameters $\omega,\aa,\bb,\cc,\di$ compatible with Eq. \eqref{eq:omabcd}, the spectrum of the reduced operator $\mathcal{A}_1$ possesses an invariant purely absolutely continuous part; in addition to this, at most two isolated eigenvalues can appear. To be more precise, from \cite[Eq. (2.13)]{ABD95} we infer\footnote{Notice that for $\bb = 0$ we have $\aa \di = 1$ (see Eq. \eqref{eq:conomabcd}), which grants $\aa + \di \neq 0$; on the other hand the constants $\Lambda_{\pm}$ are well defined and finite for any $\bb \neq 0$ (see the forthcoming Eq. \eqref{eq:defLam}).}
\begin{eqnarray*}
& \sigma(\mathcal{A}_1) = \sigma_{ac}(\mathcal{A}_1) \cup \sigma_p(\mathcal{A}_1)\,, \qquad
\sigma_{ac}(\mathcal{A}_1) = [m^2,+\infty)\,, \nonumber \\
& \sigma_{p}(\mathcal{A}_1) = \left\{\!\begin{array}{ll}
\displaystyle{\left\{m^2 - {\cc^2 \over (\aa+\di)^2}\right\} }			& \displaystyle{ \mbox{for\, $\bb = 0$,\, $\cc/(\aa+\di) \!<\! 0$}, } \vspace{0.1cm}\\
\displaystyle{ \big\{m^2 - \Lambda_{-}^2\big\} }	& \displaystyle{ \mbox{for\, $\bb \neq 0$,\, $\Lambda_{-}\! < 0\leqslant \!\Lambda_{+}$}, } \vspace{0.1cm}\\
\displaystyle{ \big\{m^2 - \Lambda_{-}^2\,,m^2 - \Lambda_{+}^2\big\} }	& \displaystyle{ \mbox{for\, $\bb \neq 0$,\, $\Lambda_{-}\! <\!\Lambda_{+} \!< \!0$}, } \vspace{0.1cm}\\
\displaystyle{ \varnothing }	& \displaystyle{ \mbox{otherwise}, }
\end{array}\right. 
\end{eqnarray*}
where
\begin{equation}\label{eq:defLam}
\Lambda_{\pm} := {\aa + \di \over 2 \bb} \pm {\sqrt{(\aa-\di)^2 + 4} \over 2 |\bb|}\;, \qquad \mbox{for\; $\bb \neq 0$}\,.
\end{equation}
From here, by a few elementary considerations we deduce that $\mathcal{A}_1$ is positive if and only if one of the following two alternatives occurs, for $m > 0$:
\begin{equation}\label{eq:conomabcd}
\mbox{$\bb = 0$\; and\; ${\cc/(\aa+\di)} > -m$} 
\quad\qquad \mbox{or} \quad\qquad 
\mbox{$\bb \neq 0$\; and\; $\Lambda_{+} \!>\! \Lambda_{-} \!>\! - m$}\,.
\end{equation}
We shall henceforth assume the parameters $\omega,\aa,\bb,\cc,\di$ to fulfil the latter Eq. \eqref{eq:conomabcd}, in addition to the conditions previously stated in Eq. \eqref{eq:omabcd}.

To proceed, let us recall that the heat kernel associated to the reduced operator $\mathcal{A}_1$ for $m = 0$ was formerly computed in \cite{ABD95}; taking into account that the addition of a mass term only produces the overall multiplicative factor $e^{-\tau m^2}$, from \cite[Eq. (3.4)]{ABD95} (see also Eqs. (2.12) and (3.2) of the cited reference) we infer
\begin{align}\label{eq:heatdel}
& e^{-\tau \mathcal{A}_1}(x_1,y_1) 
= {e^{- m^2 \tau - {|x_1 - y_1|^2 \over 4\tau}} \over \sqrt{4\pi \tau}} \\
& \qquad +  {e^{- m^2 \tau} \over \sqrt{4\pi \tau}}\, \left\{\!\!\begin{array}{ll}
\displaystyle{ L(x_1,y_1)\,e^{- {\left(|x_1| + |y_1|\right)^2 \over 4\tau}} }	&	\\
\displaystyle{ \qquad -\, {\cc \over \aa+\di}\, \big(1 + L(x_1,y_1)\big) \int_{0}^{\infty}\!\!dw\;e^{- {\cc \over \aa+\di}\,w\, - {\left(w+|x_1|+|y_1|\right)^2 \over 4\tau}} }	& \displaystyle{ \mbox{for\; $\bb = 0$},} \vspace{0.1cm}\\
\displaystyle{ \mbox{sgn}(x_1 y_1)\,e^{- {\left(|x_1| + |y_1|\right)^2 \over 4\tau}} }	&	\\
\displaystyle{ \qquad + \int_{0}^{\infty}\!\!dw \,\left( M_{+}(x_1,y_1)\, e^{-\Lambda_{+} w} -  M_{-}(x_1,y_1)\, e^{-\Lambda_{-} w} \right)\, e^{- {\left(w+|x_1|+|y_1|\right)^2 \over 4\tau}} }	& \displaystyle{ \mbox{for\; $\bb \neq 0$},}
\end{array}
\right. \nonumber
\end{align}
where we introduced the notations ($\theta(\cdot)$ is the Heaviside step function and $\mbox{sgn}(\cdot)$ is the sign function)
\begin{gather*}
L(x_1,y_1) := {\aa - \di \over \aa + \di}\; \mbox{sgn}(x_1)\, \theta(x_1 y_1) - \left(\!1 - {2\,( \Re \omega + \mbox{sgn}(x_1)\,i\, \Im \omega) \over \aa+\di}\!\right) \theta(- x_1 y_1) \,,\\
\begin{aligned}
& \displaystyle{M_{\pm}(x_1,y_1) := {\mbox{sgn}(\bb) \over \sqrt{(\aa-\di)^2 + 4}} \left[
\theta(x_1 y_1) \Big((\aa+\di)\Lambda_{\pm} - 2 \cc - {(\aa - \di)\Lambda_{\pm} \over 2}\,\mbox{sgn}(x_1)\Big)\right.} \nonumber \\
& \hspace{6.5cm} \displaystyle{\left.
- \,\theta(- x_1 y_1)\, \Lambda_{\pm} \Big( {\aa + \di \over 2} + \Re \omega \!+\!\mbox{sgn}(x_1)\,i\, \Im \omega\Big) \right]\,.}
\end{aligned}
\end{gather*}
Notice in particular that, for any $x_1 \in \mathbb{R} \setminus \{0\}$, we have
\begin{gather}
L(x_1) \equiv L(x_1,x_1) = {\aa - \di \over \aa + \di}\; \mbox{sgn}(x_1) \,, \label{eq:LMx1}\\
\displaystyle{M_{\pm}(x_1) \equiv M_{\pm}(x_1,x_1) = {\mbox{sgn}(\bb) \over \sqrt{(\aa-\di)^2 + 4}} \Big((\aa+\di)\Lambda_{\pm} - 2 \cc - {(\aa - \di)\Lambda_{\pm} \over 2}\,\mbox{sgn}(x_1)\Big)\,.} \nonumber
\end{gather}
Substituting the above expression for $e^{-\tau \mathcal{A}_1}(x_1,y_1) $ into Eq. \eqref{eq:fiuheat}, we obtain the following integral representation for the $\zeta$-regularized vacuum polarization:
\begin{align}\label{eq:fiusemi}
& \vacc \big(\hat{\phi}^{u}(t,\x)\big)^2 \vac 
= {\kappa^u \over 2\,(4\pi)^{d/2}\, \Gamma({u+1 \over 2})} \int_{0}^{\infty}\! d\tau\;\tau^{{u - d - 1 \over 2}}\, e^{- m^2 \tau} \;\times\\
& \times \left\{\!\!\begin{array}{ll}
\displaystyle{1 + L(x_1)\,e^{- {(x_1)^2 \over \tau}}\! - {\cc \over \aa+\di}\, \big(1 + L(x_1)\big)\! \int_{0}^{\infty}\!\!dw\;e^{- {\cc \over \aa+\di}\,w \,- {\left(w+2|x_1|\right)^2 \over 4\tau}} }	& \displaystyle{ \mbox{for\; $\bb = 0$},} \vspace{0.1cm}\\
\displaystyle{1 + e^{- {(x_1)^2 \over \tau}}\! + \!\int_{0}^{\infty}\!\!dw \,\left( M_{+}(x_1)\, e^{-\Lambda_{+} w} \!-\!  M_{-}(x_1)\, e^{-\Lambda_{-} w} \right)\, e^{- {\left(w+2|x_1|\right)^2 \over 4\tau}} }	& \displaystyle{ \mbox{for\; $\bb \neq 0$}.} 
\end{array}
\right. \nonumber
\end{align}

Regarding this representation, one can make considerations analogous to those reported below Eq. \eqref{eq:fiuperf}. Especially, it can be checked by direct inspection that the integrals on the right-hand side of \eqref{eq:fiusemi} are convergent and define an analytic function of $u$ in the complex strip $\Re u > d - 1$, in agreement with the general theory.

Now, let us proceed to determine the analytic continuation of the map $u \mapsto \vacc \big(\hat{\phi}^{u}(t,\x)\big)^2 \vac$. Using once more the identities \eqref{eq:idGamma} \eqref{eq:idBesselK} introduced in the previous Section \ref{sec:perfect} and making again the change of integration variable $w = 2\,|x_1|\,v$, we obtain
\begin{align}\label{eq:fiusemiAC}
& \vacc \big(\hat{\phi}^{u}(t,\x)\big)^2 \vac 
= {m^{d - 1}\, (\kappa/m)^u\, \Gamma\left({u - d + 1 \over 2}\right) \over 2^{d+1}\,\pi^{d/2}\, \Gamma({u+1 \over 2})} 
+ {2^{{u - 3d + 1 \over 2}}\,\big(\kappa |x_1|\big)^{\!u} \over \pi^{d/2}\, \Gamma({u+1 \over 2})\,|x_1|^{d - 1}}\; \times \\
& \times \left\{\!\!\begin{array}{ll}
\displaystyle{
L(x_1)\; \mathfrak{K}_{{d - 1 - u \over 2}}\big(2 m |x_1|\big) } \vspace{-0.13cm}\\
\displaystyle{
\quad -\,\big(1 + L(x_1)\big)\, {2\,\cc\, |x_1| \over \aa+\di} \!\int_{0}^{\infty}\!\!dv\;{e^{- {2\,\cc\,|x_1| \over \aa+\di}\,v} \over (v + 1)^{d - 1 - u}} \; \mathfrak{K}_{{d - 1 - u \over 2}}\big(2 m |x_1|\, (v+1)\big)\,,}	
\quad \displaystyle{ \mbox{for\; $\bb = 0$},} 
\vspace{0.3cm}\\
\displaystyle{
\mathfrak{K}_{{d - 1 - u \over 2}}\big(2 m |x_1|\big) 
+ \,2\,M_{+}(x_1)\,|x_1| \int_{0}^{\infty}\!\!dv\; {e^{-2 \Lambda_{+} |x_1|\,v} \over (v+1)^{d - 1 - u}}\; \mathfrak{K}_{{d - 1 - u \over 2}}\big(2 m |x_1|\, (v+1)\big)
}\vspace{0.03cm}\\
\displaystyle{
\quad - \,2\,M_{-}(x_1)\,|x_1| \int_{0}^{\infty}\!\!dv\; {e^{-2 \Lambda_{-} |x_1|\,v} \over (v+1)^{d - 1 - u}}\; \mathfrak{K}_{{d - 1 - u \over 2}}\big(2 m |x_1|\, (v+1)\big)\,,}	
\qquad\quad\; \displaystyle{ \mbox{for\; $\bb \neq 0$},}
\end{array}
\right. \nonumber
\end{align}
where $\mathfrak{K}_{\nu}(w) = w^{\nu} K_{\nu}(w)$ are the functions defined in Eq. \eqref{eq:KKdef}. The same considerations reported below Eq. \eqref{eq:fiuAC} apply to the present context. As a result, Eq. \eqref{eq:fiusemiAC} yields the meromorphic extension of $\vacc \big(\hat{\phi}^{u}(t,\x)\big)^2 \vac$ to the whole complex plane, with isolated simple pole singularities at 
\begin{equation*}
u = d-1-2\ell\,, \qquad \mbox{with\; $\ell \in \{0,1,2,\dots\}$}\,.
\end{equation*}

Special attention must be paid to the fact that the first addendum on the right-hand side of Eq. \eqref{eq:fiusemiAC} has a pole at $u = 0$ if the space dimension $d$ is odd. On the contrary, all other terms in Eq. \eqref{eq:fiusemiAC} are analytic at $u = 0$. Taking these facts into account, we can proceed to compute the renormalized vacuum polarization using the general prescription \eqref{eq:fi2ren}:
\begin{equation}\label{eq:firensemi}
\vacc \hat{\phi}^{2}(t,\x) \vac_{ren} = \vacc \hat{\phi}^{2} \vac_{ren}^{(free)} + \vacc \hat{\phi}^{2}(x_1) \vac_{ren}^{(plane)}\,;
\end{equation}
\begin{equation}
\vacc \hat{\phi}^{2} \vac_{ren}^{(free)} = \left\{\!\!\begin{array}{ll}
\displaystyle{ {(-1)^{{d \over 2}}\,\pi\;m^{d - 1} \over (4 \pi)^{{d + 1 \over 2}}\; \Gamma({d + 1 \over 2})} }	&	 \displaystyle{\mbox{for $d$ even}\,,} \vspace{0.1cm} \\
\displaystyle{ {(-1)^{{d-1 \over 2}}\; m^{d - 1} \over (4 \pi)^{{d + 1 \over 2}}\; \Gamma({d+1 \over 2})} \left[H_{{d-1 \over 2}} + 2\log \left({2\kappa \over m}\right) \right]}	&	 \displaystyle{\mbox{for $d$ odd}\,;}
\end{array}\right.
\end{equation}
\begin{align}\label{eq:firenpisemi}
& \vacc \hat{\phi}^{2}(x_1) \vac_{ren}^{(plane)} 
= \\
& \left\{\!\!\begin{array}{ll}
\displaystyle{
{1 \over 2^{{3d - 1 \over 2}} \pi^{{d + 1 \over 2}}\,|x_1|^{d - 1}} \Bigg[
L(x_1)\; \mathfrak{K}_{{d - 1 \over 2}}\big(2 m |x_1|\big) }\vspace{-0.13cm} \\
\displaystyle{
\qquad -\,\big(1 + L(x_1)\big)\, {2\,\cc\, |x_1| \over \aa+\di} \!\int_{0}^{\infty}\!\!dv\;{e^{- {2\,\cc\,|x_1| \over \aa+\di}\,v} \over (v + 1)^{d - 1}} \; \mathfrak{K}_{{d - 1 \over 2}}\big(2 m |x_1|\, (v+1)\big)\Bigg]\,,}	
& \displaystyle{ \mbox{for\; $\bb = 0$},} 
\vspace{0.05cm}\\
\displaystyle{{1 \over 2^{{3d - 1 \over 2}} \pi^{{d + 1 \over 2}}\,|x_1|^{d - 1}} \Bigg[
\mathfrak{K}_{{d - 1 \over 2}}\big(2 m |x_1|\big) }\vspace{-0.05cm}\\
\displaystyle{\qquad + \,2\,M_{+}(x_1)\,|x_1| \int_{0}^{\infty}\!\!dv\; {e^{-2 \Lambda_{+} |x_1|\,v} \over (v+1)^{d - 1}}\; \mathfrak{K}_{{d - 1 \over 2}}\big(2 m |x_1|\, (v+1)\big) } \\
\displaystyle{ \qquad - \,2\,M_{-}(x_1)\,|x_1| \int_{0}^{\infty}\!\!dv\; {e^{-2 \Lambda_{-} |x_1|\,v} \over (v+1)^{d - 1}}\; \mathfrak{K}_{{d - 1 \over 2}}\big(2 m |x_1|\, (v+1)\big)\Bigg]\,,}	
& \displaystyle{ \mbox{for\; $\bb \neq 0$}.}
\end{array}
\right. \nonumber
\end{align}

Before moving on, let us remark that $\vacc \hat{\phi}^{2}\vac_{ren}^{(free)}$ is exactly the same free-theory contribution arising in the case of a perfectly reflecting plane (cf. Eq. \eqref{eq:firenm}), while $\vacc \hat{\phi}^{2}(x_1) \vac_{ren}^{(plane)}$ contains all the information related to the semitransparent hyperplane $\pi$. As expected, from Eq. \eqref{eq:firenpisemi} it can be readily deduced that $\vacc \hat{\phi}^{2}(x_1) \vac_{ren}^{(plane)} = 0$ when $\bb = \cc = 0$ and $\omega = \aa = \di = 1$, namely when the quantum field is not affected by the presence of the hyperplane $\pi$.

\subsection{Asymptotics for $x_1 \to 0^{\pm}$ and $x_1 \to \pm \infty$}

In order to determine the asymptotic behaviour of the renormalized vacuum polarization $\vacc \hat{\phi}^{2}(t,\x) \vac_{ren}$ for small and large distances from the plane $\pi$, we retrace the same arguments already described in the previous subsection \ref{subsec:asy} for the case of a perfectly reflecting plane. Also in the present situation we just provide a leading order analysis, focusing primarily on the non-constant term $\vacc \hat{\phi}^{2}(x_1) \vac_{ren}^{(plane)}$ of Eq. \eqref{eq:firenpisemi}.

\subsubsection{The limit $x_1 \to 0^{\pm}$}\label{subsubsec:x0msemi}

First of all, recall the asymptotic expansions \eqref{eq:K0asy} and \eqref{eq:Knuasy} for the functions $\mathfrak{K}_{\nu}$. Taking these into account, regarding the integral expressions in Eq. \eqref{eq:firenpisemi} we infer the following for $|x_1| \to 0$ (cf. paragraph \ref{par:x10}):
\begin{align*}
& {2\,\cc\, |x_1| \over \aa+\di} \!\int_{0}^{\infty}\!\!dv\;{e^{- {2\,\cc\,|x_1| \over \aa+\di}\,v} \over (v + 1)^{d - 1}} \; \mathfrak{K}_{{d - 1 \over 2}}\big(2 m |x_1|\, (v+1)\big) \\
& = {\cc\,\big(2m |x_1|\big)^{d - 1} \over (\aa+d) m}\, e^{{2\,\cc\,|x_1| \over \aa+d}}\! \int_{2m |x_1|}^{\infty}\!\!dz\;{e^{- {\cc \over (\aa+\di) m}\,z} \over z^{d-1}} \, \mathfrak{K}_{{d - 1 \over 2}}(z)
= \left\{\!\! \begin{array}{ll}
\displaystyle{ \mathcal{O}(1) }	&	\displaystyle{\mbox{for\; $d = 1$}\,,} \vspace{0.1cm} \\
\displaystyle{ \mathcal{O}\big( |x_1|\, \log |x_1|\big) }	&	\displaystyle{\mbox{for\; $d = 2$}\,,} \vspace{0.1cm} \\
\displaystyle{ \mathcal{O}\big(|x_1|\big) }	&	\displaystyle{\mbox{for\; $d \geqslant 3$}\,;} 
\end{array} \right. \nonumber
\\
& 2\,|x_1| \int_{0}^{\infty}\!\!dv\; {e^{-2 \Lambda_{\pm} |x_1|\,v} \over (v+1)^{d - 1}}\; \mathfrak{K}_{{d - 1 \over 2}}\big(2 m |x_1|\, (v+1)\big) \\
& = {\big(2m |x_1|\big)^{d - 1} \over m}\, e^{2 \Lambda_{\pm} |x_1|} \int_{2m |x_1|}^{\infty}\!\!dz\; {e^{- {\Lambda_{\pm} \over m}\,z} \over z^{d-1}}\, \mathfrak{K}_{{d - 1 \over 2}}(z)
= \left\{\!\! \begin{array}{ll}
\displaystyle{ \mathcal{O}(1) }	&	\displaystyle{\mbox{for\; $d = 1$}\,,} \vspace{0.1cm} \\
\displaystyle{ \mathcal{O}\big( |x_1|\, \log |x_1|\big) }	&	\displaystyle{\mbox{for\; $d = 2$}\,,} \vspace{0.1cm} \\
\displaystyle{ \mathcal{O}\big(|x_1|\big) }	&	\displaystyle{\mbox{for\; $d \geqslant 3$}\,.} 
\end{array} \right. \nonumber
\end{align*}

Taking the above estimates into account, from Eq. \eqref{eq:firenpisemi} we infer for $x_1 \to 0^{\pm}$
\begin{align}\label{eq:firenx0semi}
& \vacc \hat{\phi}^{2}(x_1) \vac_{ren}^{(plane)} 
= \left\{\!\!\begin{array}{ll}
\displaystyle{-\,{\mbox{sgn}(x_1) \over 2 \pi} \left( {\aa - \di \over \aa + \di}\right) \log\big(m |x_1|\big) + \mathcal{O}(1)}
	& \displaystyle{ \mbox{for\; $d = 1$ \,and\, $\bb = 0$},} 
\vspace{0.05cm}\\
\displaystyle{-\,{1 \over 2 \pi}\,\log\big(m |x_1|\big) + \mathcal{O}(1) }	
	& \displaystyle{ \mbox{for\; $d = 1$ \,and\, $\bb \neq 0$},}
\vspace{0.1cm}\\
\displaystyle{{\mbox{sgn}(x_1) \over 8 \pi\,|x_1|} \left({\aa - \di \over \aa + \di}\right) + \mathcal{O}\big( \log |x_1|\big)}	
	& \displaystyle{ \mbox{for\; $d = 2$ \,and\, $\bb = 0$},} 
\vspace{0.1cm}\\
\displaystyle{{1 \over 8 \pi\,|x_1|} + \mathcal{O}\big(\log |x_1|\big)}	
	& \displaystyle{ \mbox{for\; $d = 2$ \,and\, $\bb \neq 0$}.}
\vspace{0.1cm}\\
\displaystyle{
{\mbox{sgn}(x_1) \,\Gamma({d - 1 \over 2}) \over (4 \pi)^{{d + 1 \over 2}}\,|x_1|^{d - 1}} \left({\aa - \di \over \aa + \di}\right) \Big[ 1 + \mathcal{O}\big(|x_1|\big) \Big]}	
	& \displaystyle{ \mbox{for\; $d \geqslant 3$ \,and\, $\bb = 0$},} 
\vspace{0.1cm}\\
\displaystyle{{\Gamma({d - 1 \over 2}) \over (4 \pi)^{{d + 1 \over 2}}\,|x_1|^{d - 1}}\, \Big[ 1 + \mathcal{O}\big(|x_1|\big) \Big]}	
	& \displaystyle{ \mbox{for\; $d \geqslant 3$ \,and\, $\bb \neq 0$}.}
\end{array}
\right. \nonumber
\end{align}

Let us briefly comment the above results. Comparing Eqs. \eqref{eq:firenx0} and \eqref{eq:firenx0semi}, it appears that $\vacc \hat{\phi}^{2}(x_1) \vac_{ren}^{(plane)}$ presents the same kind of divergence near the plane $\pi$, whether the latter be perfectly reflecting or semitransparent; as a matter of fact, the leading order terms in \eqref{eq:firenx0} and \eqref{eq:firenx0semi} exactly coincide for any $d \geqslant 1$ when $\bb \neq 0$. \\
On the other side, the expansions in Eq. \eqref{eq:firenx0semi} call the attention to two subfamilies, parametrized by
\begin{equation}\label{eq:nodiv}
\bb = 0\,, \qquad \cc \in \mathbb{R}\,, \qquad \aa = \di = \pm 1\,.
\end{equation}

In these cases the leading order contribution vanishes identically (for any $d \geqslant 1$), implying that the divergence of the renormalized vacuum polarization $\vacc \hat{\phi}^{2}(x_1) \vac_{ren}^{(plane)}$ near the hyperplane $\pi$ is somehow softened. While the occurrence of this phenomenon appears to be accidental for $\aa = \di = - 1$, some intuition can be gained instead regarding the case with $\aa = \di = + 1$.
We already mentioned that the subfamily with $\bb = 0$, $\cc \in \mathbb{R}$ and $\aa = \di = 1$ describes a delta-type potential concentrated on the hyperplane $\pi$.\footnote{\label{foot:delta}More precisely, in this case the functions belonging to the domain of the reduced operator $\mathcal{A}_1$ are continuous at $x_1 = 0$, namely $\psi(0^{+}) = \psi(0^{-}) = \psi(0)$, with discontinuous first derivative fulfilling the jump condition $\psi'(0^{+}) - \psi'(0^{-}) = \gamma\,\psi(0)$.} This is actually the ``less singular'' distributional potential amid the ones associated to the boundary conditions written in Eq. \eqref{eq:Asemi}. There are at least three interdependent ways to understand the latter claim:\\
\emph{i)} Except for the pure delta case, all distributional potentials mentioned below Eq. \eqref{eq:Asemi} comprise at least one derivative of the Dirac delta function (see \cite{Se86}, \cite[\S 3.2.4]{AK99}). It is therefore evident that, as distributions, they are more singular than the Dirac delta function itself.\\
\emph{ii)} In the case of a delta potential, the field is required to be continuous across the plane $\pi$ where the potential is concentrated (see footnote \ref{foot:delta}). In contrast, in all other cases the field exhibits a discontinuity.\\
\emph{iii)} It is well-known that a delta potential concentrated on a surface of co-dimension $1$ (such as the hyperplane $\pi$) can be approximated (in resolvent sense) by regular short-range interactions \cite[\S I.3.2]{AGHH05}. In light of this, delta potentials can be reasonably regarded as a crossing point between smooth background potentials and classical hard-wall boundaries. Given that renormalized Casimir observables present no singularity when the external potentials are smooth, it is not entirely surprising that the boundary behaviour is less singular in the case of delta potentials. The above line of thinking does not apply in the case of non-pure delta potentials, since the approximation of the latters by regular potentials is far more problematic \cite{Se87}.

Let us finally recognize that, whenever the leading order terms in the asymptotic expansions \eqref{eq:firenx0semi} vanish, the study of the sub-leading contributions becomes crucial. A detailed investigation of this subject is deferred to future works.

\subsubsection{The limit $x_1 \to \pm\infty$}

In this paragraph we proceed to examine the behaviour of the renormalized vacuum polarization in the regime of large distances from the hyperplane $\pi$. Recalling once more that the term $\vacc \hat{\phi}^{2} \vac_{ren}^{(free)}$ is constant, we restrict our analysis to the expression $\vacc \hat{\phi}^{2}(x_1) \vac_{ren}^{(plane)}$.

Firstly, recall the asymptotic expansion \eqref{eq:Kinf} for the functions $\mathfrak{K}_{\nu}$. Then, making the change of variable $z = |x_1|\, v$, we derive the following expansions of the integral expressions in Eq. \eqref{eq:firenpisemi} for $|x_1| \to +\infty$:
\begin{align*}
& {2\,\cc\, |x_1| \over \aa+\di} \!\int_{0}^{\infty}\!\!dv\;{e^{- {2\,\cc\,|x_1| \over \aa+\di}\,v} \over (v + 1)^{d - 1}} \; \mathfrak{K}_{{d - 1 \over 2}}\big(2 m |x_1|\, (v+1)\big) \\
& = {\sqrt{2\pi}\;\cc \over \aa+\di}\; e^{-2 m |x_1|}\,\big(2 m |x_1|\big)^{\!{d - 2 \over 2}} \!\int_{0}^{\infty}\!\!dz\;e^{- 2\big({\cc \over \aa+\di} + m\big) z} \; \Big(1 + \mathcal{O}\big(z/|x_1|\big) \Big)\\
& = \sqrt{{\pi \over 2}}\; {{\cc \over \aa+\di} \over {\cc \over \aa+\di} + m}\; e^{-2 m |x_1|}\,\big(2 m |x_1|\big)^{\!{d - 2 \over 2}}\, \Big(1 + \mathcal{O}\big(1/|x_1|\big) \Big)\,; \nonumber
\\
& 2\,|x_1| \int_{0}^{\infty}\!\!dv\; {e^{-2 \Lambda_{\pm} |x_1|\,v} \over (v+1)^{d - 1}}\; \mathfrak{K}_{{d - 1 \over 2}}\big(2 m |x_1|\, (v+1)\big) \\
& = \sqrt{2\pi}\, e^{- 2 m |x_1|} \big(2 m |x_1|\big)^{\!{d - 2 \over 2}}\! \int_{0}^{\infty}\!\!dz\; e^{-2 (\Lambda_{\pm} + m) z} \; \Big(1 + \mathcal{O}\big(z/|x_1|\big) \Big) \\
& = \sqrt{{\pi \over 2}}\; {1 \over \Lambda_{\pm} + m}\; e^{- 2 m |x_1|} \big(2 m |x_1|\big)^{\!{d - 2 \over 2}}\, \Big(1 + \mathcal{O}\big(1/|x_1|\big) \Big)\,. \nonumber
\end{align*}

From here and from Eq. \eqref{eq:firenpisemi}, in the limit $x_1 \to \pm \infty$ we infer
\begin{align}
& \vacc \hat{\phi}^{2}(x_1) \vac_{ren}^{(plane)} 
= {m^{{d - 2 \over 2}} \over 2 (4\pi)^{d/2}}\;{e^{-2 m |x_1|} \over |x_1|^{d/2}} \; \times \\
& \qquad \times \left\{\!\!\begin{array}{ll}
\displaystyle{ {(\aa - \di) m\; \mbox{sgn}(x_1) - \cc \over (\aa+\di)m + \cc}\, \left[1 + \mathcal{O}\left({1 \over |x_1|} \right) \right]}	
& \displaystyle{ \mbox{for\; $\bb = 0$},} 
\vspace{0.05cm}\\
\displaystyle{ 
{6 \cc + 2 \bb m^2 + 4(\aa+\di)m -(\aa-\di)m\;\mbox{sgn}(x_1) \over 2(\cc + \bb m^2 + (\aa+\di)m)}\, \left[1 + \mathcal{O}\left({1 \over |x_1|} \right) \right]}	
& \displaystyle{ \mbox{for\; $\bb \neq 0$}.}
\end{array}
\right. \nonumber
\end{align}

The above relations show that also in the case of a semitransparent plane the renormalized expectation $\vacc \hat{\phi}^{2}(x_1) \vac_{ren}^{(plane)}$ decays exponentially fast far away from the hyperplane $\pi$. In other words, the difference between the full vacuum polarization $\vacc \hat{\phi}^{2}(t,\x) \vac_{ren}$ and the constant free-theory term $\vacc \hat{\phi}^{2} \vac_{ren}^{(free)}$ becomes exponentially small.

\subsection{Vacuum polarization for a massless field}\label{subsec:m0semi}

We now examine the renormalized vacuum polarization for a massless field in presence of a semitransparent hyperplane. Making reference to Eqs. \eqref{eq:Asemi} and \eqref{eq:conomabcd}, for a sensible quantum field theory we must require either
\begin{equation}
\mbox{$\bb = 0$\; and\; ${\cc/(\aa+\di)} \geqslant 0$}
\quad\qquad \mbox{or} \quad\qquad 
\mbox{$\bb \neq 0$\; and\; $\Lambda_{+} \!>\! \Lambda_{-} \!\geqslant\! 0$}\,.
\end{equation}

In accordance with the general arguments of Section \ref{sec:gen} (see, especially, Eq. \eqref{eq:fi2renm0}), we proceed to determine the renormalized observable of interest evaluating the zero-mass limit $m \to 0^{+}$ of the analogous quantity in the massive theory.

Similarly to the configuration with a perfectly reflecting surface, the cases with space dimension $d = 1$ and $d \geqslant 2$ need to be analysed separately.

\subsubsection{Space dimension $d = 1$}\label{subsubsec:m0d1semi}

A careful analysis is demanded for this specific model, due to the emergence of the same infrared pathologies already discussed in paragraph \ref{subsubsec:m0d1}. Indeed, let us recall that the renormalized vacuum polarization comprises a free theory contribution which is divergent in the limit $m \to 0^{+}$ (see Eq. \eqref{eq:firenm1d}):
\begin{equation*}
\vacc \hat{\phi}^{2}\vac_{ren}^{(free)} = {1 \over 2 \pi}\, \log \left({2\kappa \over m}\right) .
\end{equation*}

On the other hand, by arguments similar to those described in paragraph \ref{subsubsec:m0d1}, from Eq. \eqref{eq:firenpi} we obtain the following asymptotic expansions for $m \to 0^{+}$, involving the incomplete Gamma function $\Gamma(a,z)$:\footnote{Notice also the following basic identity, which can be easily deduced from Eqs. \eqref{eq:defLam} and \eqref{eq:LMx1}:
$$
1 + {M_{+}(x_1) \over \Lambda_{+}} - {M_{-}(x_1) \over \Lambda_{-}} = 3\,.
$$
}
\begin{align}
& \vacc \hat{\phi}^{2}(x_1) \vac_{ren}^{(plane)}
= \left\{\!\!\begin{array}{ll}
\displaystyle{
{1 \over 2 \pi} \Bigg[
\log\big(m |x_1|\big) - \gamma_{EM} 
+ \big(1 + L(x_1)\big)\, e^{{2\,\cc\, |x_1| \over \aa+\di}}\,\Gamma\left(0\,,{2\,\cc\, |x_1| \over \aa + \di}\right)
\Bigg]} \\
\displaystyle{\qquad +\; \mathcal{O}\Big(\big(m |x_1|\big)^2 \log \big(m |x_1|\big) \Big)}	
\hspace{5.5cm} \displaystyle{ \mbox{for\; $\bb = 0$},} 
\vspace{0.05cm}\\
\displaystyle{{1 \over 2 \pi} \Bigg[
- 3\, \log\big(m |x_1|\big) + 3\,\gamma_{EM} }\vspace{-0.05cm}\\
\displaystyle{\qquad  
-\, {M_{+}(x_1) \over \Lambda_{+}}\; e^{2 \Lambda_{+} |x_1|}\, \Gamma\big(0,2 \Lambda_{+} |x_1|\big)
+ {M_{-}(x_1) \over \Lambda_{-}}\; e^{2 \Lambda_{-} |x_1|}\, \Gamma\big(0,2 \Lambda_{-} |x_1|\big)\Bigg]} \\
\displaystyle{\qquad +\; \mathcal{O}\Big(\big(m |x_1|\big)^2 \log \big(m |x_1|\big) \Big)}	
\hspace{5.5cm} \displaystyle{ \mbox{for\; $\bb \neq 0$}.}
\end{array}
\right. \nonumber
\end{align}

For $\bb \neq 0$, the above relations together with Eq. \eqref{eq:firensemi} make evident that
\begin{equation*}
\lim_{m \to 0^{+}}\vacc \hat{\phi}^{2}(t,\x) \vac_{ren} 
= \lim_{m \to 0^{+}} \left[{1 \over 2 \pi} \log \left({2\kappa \over m^4 |x_1|^3}\right) + \mathcal{O}(1)\right] = +\infty,
\end{equation*}
indicating that in this case the renormalized expectation $\vacc \hat{\phi}^{2}(t,\x) \vac_{ren}^{massless}$ is irremediably divergent in the infrared for a massless field. Before we proceed, let us point out a connection between this result and the similar conclusions drawn in paragraph \ref{subsubsec:m0d1} for the case of Neumann conditions on the hyperplane $\pi$: Neumann conditions are formally recovered in the present scenario as the limit case where $\bb \to \infty$, with $\cc = 0$, $\omega = \aa = \di = 1$.

Let us henceforth assume
\begin{equation*}
\bb = 0\,.
\end{equation*}
Under this condition, from the above results and from Eq. \eqref{eq:fi2renm0} we readily infer
\begin{align}
& \vacc \hat{\phi}^{2}(t,\x) \vac_{ren}^{(massless)} 
= \lim_{m \to 0^{+}} \Big[ \vacc \hat{\phi}^{2} \vac_{ren}^{(free)} + \vacc \hat{\phi}^{2}(x_1) \vac_{ren}^{(plane)} \Big] \nonumber \\
& = {1 \over 2 \pi} \left[\log\!\big(2\kappa |x_1|\big) - \gamma_{EM} 
+ \left(1 + {\aa - \di \over \aa + \di}\; \mbox{sgn}(x_1)\right)\, e^{{2\,\cc\, |x_1| \over \aa+\di}}\,\Gamma\left(0\,,{2\,\cc\, |x_1| \over \aa + \di}\right)\right]\,.
\end{align}

Using again known properties of the incomplete Gamma function, it is easy to derive the following leading order asymptotic expansions of the the above expression:
\begin{equation*}
\vacc \hat{\phi}^{2}(t,\x) \vac_{ren}^{(massless)} = \left\{\!\!\begin{array}{ll}
\displaystyle{ \mp\,{1 \over 2 \pi} \left({\aa - \di \over \aa + \di}\right) \log\left({\cc\, |x_1| \over \aa + \di}\right) + \mathcal{O}(1) }	&	\displaystyle{\mbox{for\, $x_1 \to 0^{\pm}$},} \vspace{0.1cm} \\
\displaystyle{ {1 \over 2 \pi}\,\log\big(\kappa |x_1|\big)  + \mathcal{O}(1) }	&	\displaystyle{\mbox{for\, $x_1 \to \pm \infty$}\,.}
\end{array}
\right.
\end{equation*}

It is remarkable that the renormalized vacuum polarization $\vacc \hat{\phi}^{2}(t,\x) \vac_{ren}^{(massless)}$ remains finite in the limit $x_1 \to 0^{\pm}$ when $\bb = 0$, $\cc \in \mathbb{R}$ and $\aa = \di = \pm 1$. Recall that the very same phenomenon occurs in the case of a massive field (see paragraph \ref{subsubsec:x0msemi} and the comments reported therein).

\subsubsection{Space dimension $d \geqslant 2$}

Recall that for $d \geqslant 2$ the free-theory contribution $\vacc \hat{\phi}^{2} \vac_{ren}^{(free)}$ vanishes in the limit $m \to 0^{+}$ (see Eq. \eqref{eq:firenm}). Taking this into account, by arguments analogous to those described in paragraph \ref{subsubsec:d2perfm0}, from Eq. \eqref{eq:firenpisemi} we get
\begin{align}
& \vacc \hat{\phi}^{2}(t,\x) \vac_{ren}^{(massless)} = \lim_{m \to 0^{+}} \vacc \hat{\phi}^{2}(x_1) \vac_{ren}^{(plane)} \\
& = \left\{\!\!\begin{array}{ll}
\displaystyle{
{\Gamma({d - 1 \over 2}) \over (4\pi)^{{d + 1 \over 2}}\,|x_1|^{d - 1}} \left[
L(x_1) -\,\big(1 + L(x_1)\big)\,e^{{2\,\cc\,|x_1| \over \aa+\di}} \left({2\,\cc\,|x_1| \over \aa+\di}\right)^{\!d-1} \Gamma\left(2-d,{2\,\cc\, |x_1| \over \aa+\di}\right)\right]} \\
\hspace{8cm} \displaystyle{ \mbox{for\; $\bb = 0$},} 
\vspace{0.05cm}\\
\displaystyle{{\Gamma({d - 1 \over 2}) \over (4 \pi)^{{d + 1 \over 2}}\,|x_1|^{d - 1}} \left[
1 + {M_{+}(x_1) \over \Lambda_{+}}\;e^{2 \Lambda_{+} |x_1|}\;\big(2 \Lambda_{+}\,|x_1|\big)^{d-1}\, \Gamma\big(2-d,2 \Lambda_{+} |x_1|\big) \right.} \\
\displaystyle{ \hspace{3cm} \left. - \,{M_{-}(x_1) \over \Lambda_{-}}\;e^{2 \Lambda_{-} |x_1|}\; \big(2 \Lambda_{-}\,|x_1|\big)^{d-1}\, \Gamma\big(2-d,2 \Lambda_{-} |x_1|\big)\right]} \\
\hspace{8cm} \displaystyle{ \mbox{for\; $\bb \neq 0$}.}
\end{array}
\right. \nonumber
\end{align}

Using once more the known expansions of the incomplete Gamma function for small and large values of the argument, we derive the following leading order asymptotics:
\begin{align}
& \vacc \hat{\phi}^{2}(t,\x) \vac_{ren}^{(massless)} 
= \left\{\!\!\begin{array}{ll}
\displaystyle{ \pm \left({\aa - \di \over \aa + \di}\right) {1 \over 8\pi\,|x_1|} + \mathcal{O}\big(\log|x_1|\big)}	
& \displaystyle{ \mbox{for\; $\bb = 0$, $d = 2$ and $x_1 \to 0^{\pm}$},}  
\vspace{0.1cm} \\
\displaystyle{ \pm \left({\aa - \di \over \aa + \di}\right) {\Gamma({d - 1 \over 2}) \over (4\pi)^{{d + 1 \over 2}}\,|x_1|^{d - 1}}\, \Big[1 + \mathcal{O}\big(|x_1| \big)\Big]}	
& \displaystyle{ \mbox{for\; $\bb = 0$, $d \geqslant 3$ and $x_1 \to 0^{\pm}$},} 
\vspace{0.1cm}\\
\displaystyle{ -\,{\Gamma({d - 1 \over 2}) \over (4\pi)^{{d + 1 \over 2}}\,|x_1|^{d - 1}}\,\Big[1 + \mathcal{O}\big(1/|x_1| \big) \Big]} & 
\displaystyle{ \mbox{for\; $\bb = 0$, $d \geqslant 2$ and $x_1 \to \pm \infty$};} 
\vspace{0.1cm}\\
\displaystyle{{1 \over 8\pi\,|x_1|} + \mathcal{O}\big(\log|x_1|\big)}	
& \displaystyle{ \mbox{for\; $\bb \neq 0$, $d = 2$ and $x_1 \to 0^{\pm}$},} 
\vspace{0.1cm} \\
\displaystyle{{\Gamma({d - 1 \over 2}) \over (4 \pi)^{{d + 1 \over 2}}\,|x_1|^{d - 1}}\, \Big[1 + \mathcal{O}\big(|x_1| \big)\Big]}	
& \displaystyle{ \mbox{for\; $\bb \neq 0$, $d \geqslant 3$ and $x_1 \to 0^{\pm}$},} 
\vspace{0.1cm} \\
\displaystyle{{3\,\Gamma({d - 1 \over 2}) \over (4 \pi)^{{d + 1 \over 2}}\,|x_1|^{d - 1}}\,\Big[1 + \mathcal{O}\big(1/|x_1| \big) \Big]} 
& \displaystyle{ \mbox{for\; $\bb \neq 0$, $d \geqslant 2$ and $x_1 \to \pm \infty$}.}
\end{array}
\right. \nonumber
\end{align}

Notice that also in this case the local divergences of the renormalized polarization $\vacc \hat{\phi}^{2}(t,\x) \vac_{ren}^{(massless)}$ near the hyperplane $\pi$ are softened for $\bb = 0$, $\cc \in \mathbb{R}$ and $a = d = \pm 1$; again, we refer to the analysis of paragraph \ref{subsubsec:x0msemi}.
\vspace{0.5cm}

\textbf{Aknowledgments.} I am grateful to Livio Pizzocchero for many interesting conversations, some of which inspired the subject of this work. I also wish to thank Claudio Cacciapuoti for precious insights on the representation formulae for the heat kernel.
\vspace{0.2cm}

\textbf{Funding.} This work was partly supported by Progetto Giovani INdAM-GNFM 2020 ``\textit{Emergent Features in Quantum Bosonic Theories and Semiclassical Analysis}'' (Istituto Nazionale di Alta Matematica `Francesco Severi' - Gruppo Nazionale per la Fisica Matematica).

\appendix

\section{The heat kernel for a perfectly reflecting plane}\label{app:heatperf}
In this appendix we report more details about the derivation of Eq. \eqref{eq:heatrob}, regarding the heat kernel $e^{-\tau \mathcal{A}_1}(x_1,y_1)$ associated to the operator $\mathcal{A}_1$ defined in Eq. \eqref{eq:Aperf}. For the sake of presentation, hereafter we restrict the attention to the massless case, fixing
\begin{equation}\label{eq:m0}
m = 0\,.
\end{equation}
Of course this implies no loss of generality, since the heat kernel for $m > 0$ can always be recovered by an elementary shift:
\begin{equation}\label{eq:heatm0}
e^{-\tau \mathcal{A}_1}(x_1,y_1)\Big|_{m \,>\, 0} = e^{- m^2 \tau} \Big(e^{-\tau \mathcal{A}_1}(x_1,y_1)\Big|_{m \,=\, 0} \Big)\,.
\end{equation}
As far as the present analysis is concerned, it is further convenient to refer to the decomposition $L^2(\mathbb{R}) \equiv L^2(\mathbb{R}_{+}) \oplus L^2(\mathbb{R}_{-})$ (where $\mathbb{R}_{+} \equiv (0,+\infty)$ and $\mathbb{R}_{-} \equiv (-\infty,0)$) and to consider the equivalent characterization of $\mathcal{A}_1$ given by (cf. \eqref{eq:Aperf} and \eqref{eq:bcb})
\begin{eqnarray}
& \mathcal{A}_1 = A_{+} \!\oplus A_{-}\,, \nonumber \\
& \mbox{dom}(A_{\pm}) := \big\{\psi_{\pm} \in H^2(\mathbb{R}_{\pm})\,\big|\, \mp \psi_{\pm}'(0^{\pm}) + \gg_{\pm}\, \psi_{\pm}(0^{\pm}) = 0\big\}\,, \nonumber \\
& A_{\pm} \psi_{\pm} = - \,\psi''_{\pm} \quad \mbox{in\; $\mathbb{R}_{\pm}$}\,. \label{eq:Apmdom}
\end{eqnarray}

Against this background, the heat kernel can be expressed as
\begin{equation}\label{eq:heatA1Apm}
e^{-\tau \mathcal{A}_1}(x_1,y_1) = \left\{\!\begin{array}{ll}
\displaystyle{e^{-\tau A_{+}}(x_1,y_1)} & \mbox{for\; $x_1,y_1 > 0$}\,, \\
\displaystyle{e^{-\tau A_{-}}(x_1,y_1)} & \mbox{for\; $x_1,y_1 < 0$}\,.
\end{array}\right.
\end{equation}
Furthermore, the integral kernels $e^{-\tau A_{+}}(x_1,y_1)$ on $\mathbb{R}_{+}$ and $e^{-\tau A_{-}}(x_1,y_1)$ on $\mathbb{R}_{-}$ are related as follows by an elementary reflection argument:
\begin{equation}\label{eq:heatAmAp}
e^{-\tau A_{-}}(x_1,y_1) = e^{-\tau A_{+}}(-x_1,-y_1)\Big|_{\gg_{+} =\, \gg_{-}} \qquad \mbox{for\; $x_1,y_1 < 0$}\,.
\end{equation}

Before proceeding, let us point out that an explicit expression for the heat kernel on the half-line with Robin boundary conditions, namely $e^{-\tau A_{+}}(x_1,y_1)$, was previously derived in \cite{BF05} for $\gg_{+} > 0$, via an elegant technique allowing to translate Robin problems into Dirichlet ones and \textit{vice versa}. However, a few variations of this approach are required when $\gg_{+} < 0$, in order to include the discrete spectrum contribution arising in this case.

Here we prefer to present a computation of $e^{-\tau A_{+}}(x_1,y_1)$ for any $\gg_{+} \in \mathbb{R}$ (including $\gg = +\infty$ as a limit case), starting from its eigenfunction expansion (an approach in fact also hinted at in \cite{BF05}). To this purpose, we firstly notice that the spectrum of the operator $A_{+}$ defined in Eq. \eqref{eq:Apmdom} is
\begin{eqnarray*}
& \sigma(A_{+}) = \sigma_{ac}(A_{+}) \cup \sigma_{p}(A_{+})\,; \nonumber \\
& \sigma_{ac}(A_{+}) = [0,+\infty)\,, \qquad 
\sigma_{p}(A_{+}) = \left\{\!\!\begin{array}{ll}
\displaystyle{ \varnothing }		&	\displaystyle{\mbox{for\; $\gg_{+} \in [0,+\infty) \cup \{+\infty\}$}}\,, \\
\displaystyle{ \{-\gg_{+}^2\} }	&	\displaystyle{\mbox{for\; $\gg_{+} \in (-\infty,0)$}}\,.
\end{array}\right.
\end{eqnarray*}

Accordingly, we refer to the Hilbert space decomposition $L^2(\mathbb{R}_{+}) = L^2_{ac}(\mathbb{R}_{+}) \oplus L^2_{p}(\mathbb{R}_{+})$, involving the subspace $L^2_{ac}(\mathbb{R}_{+})$ of absolute continuity for $A_{+}$ and its orthogonal complement $L^2_{p}(\mathbb{R}_{+})$. A complete set of generalized eigenfunctions $\{\psi_k\}_{k \in (0,+\infty)}$ spanning $L^2_{ac}(\mathbb{R}_{+})$ and the normalized eigenfunction $\psi_{\gg_{+}} \!\in\! L^2(\mathbb{R}_{+})$ associated to the possible negative eigenvalue are given by \cite[Eq. (3.15)]{BF05}
\begin{eqnarray*}
& \displaystyle{\psi_k(x_1) := \sqrt{2 \over \pi}\;{1 \over \sqrt{k^2 \!+ \gg_{+}^2}}\, \Big(k\,\cos(k x_1) + \gg_{+} \sin(k x_1) \Big) \qquad (k > 0)\,;} \\
& \displaystyle{\psi_{\gg_{+}}(x_1) := \sqrt{2\, |\gg_{+}|}\;e^{-\,|\gg_{+}|\, x_1} \qquad \mbox{for\; $\gg_{+} < 0$}\,.}
\end{eqnarray*}

In light of the above considerations, we obtain the eigenfunction expansion (for $\tau > 0$ and $x_1,y_1 > 0$)
\begin{align*}
& \hspace{-0.cm} e^{-\tau A_{+}}(x_1,y_1) = 
\int_{0}^{\infty}\! dk\;e^{-\tau\, k^2}\,\psi_k(x_1)\, \psi_k(y_1)
+ \theta(-\gg_{+})\, e^{-\tau\, (-\gg_{+}^2)}\, \psi_{\gg_{+}}\!(x_1)\, \psi_{\gg_{+}}\!(y_1) \nonumber \\
& = {2 \over \pi} \int_{0}^{\infty}\! dk\;e^{-\tau\, k^2} {1 \over k^2 \!+ \gg_{+}^2}\, \Big(k\,\cos(k x_1) + \gg_{+} \sin(k x_1) \Big) \Big(k\,\cos(k y_1) + \gg_{+} \sin(k y_1) \Big)
\nonumber \\
& \qquad + \theta(-\gg_{+})\,2\, |\gg_{+}|\; e^{\tau\, \gg_{+}^2 -\,|\gg_{+}|\, (x_1 + y_1)}\,. \nonumber 
\end{align*}
By simple trigonometric identities and using \cite[Eqs. 3.954]{GR}, from here it follows
\begin{align*}
& e^{-\tau A_{+}}(x_1,y_1) = {1 \over \pi} \int_{0}^{\infty}\! dk\;e^{-\tau\, k^2} 
\Bigg[
\cos\big(k (x_1 - y_1)\big) + \cos\big(k (x_1 + y_1)\big) \\
& \hspace{4.5cm}
- {2\gg_{+}^2 \over k^2 \!+ \gg_{+}^2}\, \cos\big(k (x_1 + y_1)\big)
+ {2\,k\, \gg_{+} \over k^2 \!+ \gg_{+}^2} \sin\big(k (x_1 + y_1)\big)
\Bigg]
\nonumber \\
& \hspace{2.5cm} + \theta(-\gg_{+})\,2\, |\gg_{+}|\; e^{\tau\, \gg_{+}^2 -\,|\gg_{+}|\, (x_1 + y_1)}\,. \nonumber \\
& = 
{1 \over \sqrt{4\pi \tau}}\; e^{- {|x_1 - y_1|^2 \over 4\tau}}
+ {1 \over \sqrt{4\pi \tau}}\; e^{- {|x_1 + y_1|^2 \over 4\tau}} 
+ \theta(-\gg_{+})\,2\, |\gg_{+}|\; e^{\tau\, \gg_{+}^2 -\,|\gg_{+}|\, (x_1 + y_1)}\\
& \qquad
- {|\gg_{+}| \over 2}\,e^{\tau \gg_{+}^2} \left[
2\,\cosh\big(|\gg_{+}| (x_1 + y_1)\big)
- e^{-|\gg_{+}| (x_1 + y_1)}\, \mbox{erf}\Big(|\gg_{+}| \sqrt{\tau} - {x_1 + y_1 \over 2 \sqrt{\tau}}\Big) \right.\\
& \hspace{7cm} \left. -\, e^{|\gg_{+}| (x_1 + y_1)}\, \mbox{erf}\Big(|\gg_{+}| \sqrt{\tau} + {x_1 + y_1 \over 2 \sqrt{\tau}}\Big)
\right]
\\
& \qquad
- {\gg_{+} \over 2}\, e^{\tau \gg_{+}^2} \left[
2\,\sinh\big(|\gg_{+}| (x_1 + y_1)\big)
+ e^{-|\gg_{+}| (x_1 + y_1)}\, \mbox{erf}\Big(|\gg_{+}| \sqrt{\tau} - {x_1 + y_1 \over 2 \sqrt{\tau}}\Big) \right.\\
& \hspace{7cm} \left.
- \, e^{|\gg_{+}| (x_1 + y_1)}\, \mbox{erf}\Big(|\gg_{+}| \sqrt{\tau} + {x_1 + y_1 \over 2 \sqrt{\tau}}\Big)
\right] . 
\end{align*}
where $\mbox{erf}(\cdot)$ indicates the usual error function \cite[\S 7]{NIST}. Performing a few elementary manipulations and using the basic identity $\mbox{erf}(-z) = - \mbox{erf}(z)$ \cite[Eq. 7.4.1]{NIST}, we obtain
\begin{align*}
e^{-\tau A_{+}}(x_1,y_1) 
& = {1 \over \sqrt{4\pi \tau}}\; e^{- {|x_1 - y_1|^2 \over 4\tau}}
+ {1 \over \sqrt{4\pi \tau}}\; e^{- {|x_1 + y_1|^2 \over 4\tau}}
- \gg_{+}\, e^{\tau\, \gg_{+}^2 + \gg_{+} (x_1 + y_1)} 
\\
& \qquad
+ \gg_{+}\,e^{\tau \gg_{+}^2} \left[
-\,\theta(-\gg_{+})\,  e^{-|\gg_{+}| (x_1 + y_1)}\, \mbox{erf}\Big(|\gg_{+}| \sqrt{\tau} - {x_1 + y_1 \over 2 \sqrt{\tau}}\Big) \right.\\
& \hspace{3.5cm} \left. +\, \theta(\gg_{+})\, e^{|\gg_{+}| (x_1 + y_1)}\, \mbox{erf}\Big(|\gg_{+}| \sqrt{\tau} + {x_1 + y_1 \over 2 \sqrt{\tau}}\Big)
\right] \\
& = {1 \over \sqrt{4\pi \tau}}\; e^{- {|x_1 - y_1|^2 \over 4\tau}}
+ {1 \over \sqrt{4\pi \tau}}\; e^{- {|x_1 + y_1|^2 \over 4\tau}}
\\
& \hspace{3cm}
- \gg_{+}\,e^{\tau \gg_{+}^2 + \gg_{+} (x_1 + y_1)}  \left[
1 -  \mbox{erf}\Big(\gg_{+} \sqrt{\tau} + {x_1 + y_1 \over 2 \sqrt{\tau}}\Big) \right].
\end{align*}

On account of \cite[Eq. 7.2.2]{NIST} and via a basic change of the integration variable, from the latter result we get
\begin{align}\label{eq:heatAp}
e^{-\tau A_{+}}(x_1,y_1) 
& = {1 \over \sqrt{4\pi \tau}}\; e^{- {|x_1 - y_1|^2 \over 4\tau}}
+ {1 \over \sqrt{4\pi \tau}}\; e^{- {|x_1 + y_1|^2 \over 4\tau}}
\\
& \hspace{3cm}
- {2\,\gg_{+} \over \sqrt{\pi}}\,e^{\tau \gg_{+}^2 + \gg_{+} (x_1 + y_1)} \! \int_{\gg_{+} \sqrt{\tau} + {x_1 + y_1 \over 2 \sqrt{\tau}}}^{\infty} dz\;e^{-z^2} \nonumber\\
& = {1 \over \sqrt{4\pi \tau}}\; e^{- {|x_1 - y_1|^2 \over 4\tau}}
+ {1 \over \sqrt{4\pi \tau}}\; e^{- {|x_1 + y_1|^2 \over 4\tau}}
- {\gg_{+} \over \sqrt{\pi \tau}} \int_{0}^{\infty}\! dw\; e^{- \gg_{+} w \,- {(w + x_1 + y_1)^2 \over 4 \tau}}.\nonumber
\end{align}

Together with the preceding Eqs. \eqref{eq:heatm0} \eqref{eq:heatA1Apm} and \eqref{eq:heatAmAp}, this suffices to prove Eq. \eqref{eq:heatrob} in the main text.

Let us finally comment on the limit case $\gg_{+} = +\infty$. Setting $p := \gg_{+} w$ (with $\gg_{+} > 0$ finite) in the final identity of Eq. \eqref{eq:heatAp}, we get
\begin{equation*}
e^{-\tau A_{+}}(x_1,y_1) = {1 \over \sqrt{4\pi \tau}}\; e^{- {|x_1 - y_1|^2 \over 4\tau}}
+ {1 \over \sqrt{4\pi \tau}}\; e^{- {|x_1 + y_1|^2 \over 4\tau}}
- {1 \over \sqrt{\pi \tau}} \int_{0}^{\infty}\! dp\; e^{- p \,- {(p/\gg_{+} + x_1 + y_1)^2 \over 4 \tau}}. 
\end{equation*}
Starting from here, by the dominated convergence theorem we obtain
\begin{align*}
\lim_{\gg_{+} \to +\infty} e^{-\tau A_{+}}(x_1,y_1)
& = {1 \over \sqrt{4\pi \tau}}\; e^{- {|x_1 - y_1|^2 \over 4\tau}}
+ {1 \over \sqrt{4\pi \tau}}\; e^{- {|x_1 + y_1|^2 \over 4\tau}}
- {2 \over \sqrt{4\pi \tau}}\; e^{- {|x_1 + y_1|^2 \over 4 \tau}} \int_{0}^{\infty}\! dp\; e^{- p}\\
& = {1 \over \sqrt{4\pi \tau}}\; e^{- {|x_1 - y_1|^2 \over 4\tau}}
- {1 \over \sqrt{4\pi \tau}}\; e^{- {|x_1 + y_1|^2 \over 4\tau}} , 
\end{align*}
which reproduces as expected the heat kernel on the half-line for Dirichlet boundary conditions (cf. the corresponding considerations reported in Section \ref{sec:perfect}).

\end{document}